\documentclass{article}
\usepackage[height=8.2in,width=5.2in]{geometry}

\usepackage{graphicx}

\usepackage{floatrow} 
\newfloatcommand{capbtabbox}{table}[][\FBwidth]

\usepackage{tikz}

\usepackage[sort,numbers]{natbib}

\usepackage{pgfplots}
\pgfplotsset{compat=newest}% use newest version

\usepgfplotslibrary{external}
\usepackage{amsmath}
\usepackage{amsthm}
\usepackage{amssymb} 
\usepackage{subfigure}

\usepackage{enumerate}
\usepackage{array}
\usepackage[colorlinks=true,citecolor=blue,breaklinks]{hyperref}

\usepackage{appendix}

\newtheorem{theorem}{Theorem}
\newtheorem{lemma}[theorem]{Lemma}
\newtheorem{prop}[theorem]{Proposition}
\newtheorem{corr}[theorem]{Corollary}
\theoremstyle{definition}

\newtheorem{rmk}[theorem]{Remark}

\usepackage{bm}
\newcommand{\msigma}  {\bm{\Sigma}}
\newcommand{\mgamma}  {\bm{\Gamma}}
\newcommand{\mlambda}  {\bm{\Lambda}}
\newcommand{\mq}{\bm{Q}}
\newcommand{\mn}{\bm{N}}
\newcommand{\md}{\bm{D}}
\newcommand{\mm}{\bm{M}} 
\newcommand{\mx}{\bm{X}}       
\newcommand{\my}{\bm{Y}}       
\newcommand{\mi}{\bm{I}}       
\newcommand{\mU}{\bm{U}}       
\newcommand{\mb}{\bm{B}}       
\newcommand{\ms}{\bm{S}}

\newcommand{\vtheta}  {\bm{\theta}}

\newcommand{\vm}{\bm{m}}
\newcommand{\vt}{\bm{t}}
\newcommand{\vx}{\bm{x}}
\newcommand{\vy}{\bm{y}}
\newcommand{\vz}{\bm{z}}
\newcommand{\vu}{\bm{u}}
\newcommand{\vv}{\bm{v}}

\newcommand{\Dc}{\mathcal{D}}
\newcommand{\Gc}{\mathcal{G}}

\newcommand{\pnorm}[2]{\| {#1} \|_{#2}}

\newcommand{\enorm}[1]{\pnorm{#1}{2}}
\newcommand{\E}{\mathbb{E}}
\newcommand{\set}[1]{\{ #1\}}
\newcommand{\reals}{\mathbb{R}}
\newcommand{\nlsum}{\sum\nolimits}

\newcommand{\peg}{p_{\text{eg}}}

\DeclareMathOperator{\Diag}{Diag}

\numberwithin{equation}{section}

\newcommand{\MIT}{Laboratory for Information and Decision Systems, Massachusetts Institute of Technology, Cambridge, USA}
\newcommand{\tehr}{School of ECE, College of Engineering, University of Tehran, Tehran, Iran}
\newcommand{\wern}{Werner Reichardt Centre for Integrative Neuroscience, T\"ubingen, Germany}

\begin{document}

\title{Inference and Mixture Modeling with the \\ Elliptical Gamma Distribution}
\author{Reshad Hosseini\thanks{\tehr}
\and 
Suvrit Sra\thanks{\MIT}
\and
Lucas Theis\thanks{\wern}
\and
Matthias Bethge\thanks{\wern}}
\maketitle

\begin{abstract}
  We study modeling and inference with the \emph{Elliptical Gamma Distribution} (EGD). We consider maximum likelihood (ML) estimation for EGD scatter matrices, a task for which we develop new fixed-point algorithms. Our algorithms are efficient and converge to global optima despite nonconvexity. Moreover, they turn out to be much faster than both a well-known iterative algorithm of Kent \& Tyler (1991) and sophisticated manifold optimization algorithms. Subsequently, we invoke our ML algorithms as subroutines for estimating parameters of a mixture of EGDs. We illustrate our methods by applying them to model natural image statistics---the proposed EGD mixture model yields the most parsimonious model among several competing approaches.
\end{abstract}

\section{Introduction}
Non-Gaussian distributions occur in a multitude of applications. They may capture manifold structure~\citep{vmf,pennec,chikuse}, or elicit sparsity~\citep{kotz.laplace,seeger.sparse}, express heavy or light tailed behavior~\citep{ollila11,kotz.laplace}, characterize independence~\citep{lee99,ica}, or help us model a variety of other properties of data.

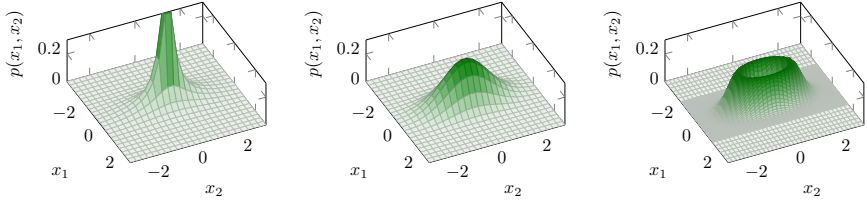
\begin{figure}[tbp]
\begin{tabular}{ccc}
\hskip-15pt
 \resizebox{.3\textwidth}{!}{% \pgfplotsset{
%   colormap={bluegreen}{color(0cm)=(blue!50!green!70); color(1cm)=(blue!50!green!50!)}
% }
\pgfplotsset{
%  colormap={cool}{rgb255(0cm)=(220,220,220); rgb255(1cm)=(0,128,255); rgb255(2cm)=(255,0,255)}
  colormap={yg}{rgb255(0cm)=(232,240,232); rgb255(1cm)=(0,128,0)}
}

\begin{tikzpicture}[scale=0.75,
    declare function={mu1=0;},
    declare function={mu2=0;},
    declare function={a=1/3;},
    declare function={b(\a)=2/\a;},
    %declare function={gamma(\z)= 2.506628274631*sqrt(1/\z)+ 0.20888568*(1/\z)^(1.5)+ 0.00870357*(1/\z)^(2.5)- (174.2106599*(1/\z)^(3.5))/25920- (715.6423511*(1/\z)^(4.5))/1244160)*exp((-ln(1/\z)-1)*\z;},
    % Following from numerical recipies -- very accurate and stable
    declare function={gammaln(\z)= ln(2.506628274631 * (
    1.00000000019001 + 76.1800917294715/\z - 86.5053203294168/(\z + 1) + 24.0140982408309/(\z + 2) - 1.23173957245016/(\z + 3) + 1.20865097386618*0.001/(\z + 4) - 5.395239384953*0.000001/(\z + 5) 
    )) - \z - 4.5 + (\z - 0.5) * ln(\z + 4.5);},
    declare function={sigma1=0.5;},
    declare function={sigma2=1;},
    declare function={normal(\m,\s)=1/(2*\s*sqrt(pi))*exp(-(x-\m)^2/(2*\s^2));},
    declare function={bivar(\ma,\sa,\mb,\sb,\aa)=
exp ( (\aa-1) *  ln((x-\ma)^2/\sa^2 + (y-\mb)^2/\sb^2 + 0.001) - (\aa/2) *((x-\ma)^2/\sa^2 + (y-\mb)^2/\sb^2) - ln(pi*\sa*\sb) - gammaln(\aa) - \aa*ln(2) + \aa*ln(\aa) );}]
%* gamma(1)/gamma(\aa)/b(\aa)^(\aa) 
\begin{axis}[
    width=5.5cm,
    xlabel=$x_1$,
    ylabel=$x_2$,
    zlabel={$p(x_1,x_2)$},
    zmin=0,
    zmax=0.301,
    samples=24, %34
    domain=-3:3,
    y domain=-3:3,
    view={65}{65},
    %colormap name=hot2,
    %plot box ratio={1 1 5},
    %height=5cm,    
    %enlargelimits=false,
    %grid=major,
   %enlarge z limits
    %colorbar,
    %colorbar style={
    %    at={(1,0)},
    %    anchor=south west,
    %    height=0.25*\pgfkeysvalueof{/pgfplots/parent axis height},
    %    title={$P(x_1,x_2)$}
    %}
]
%\addplot3[surf,samples=30,shader=interp,domain=-3:3,colormap name=whitered] {bivar(mu1,sigma1,mu2,sigma2,a)}; 
%\addplot3[mesh,samples=7] {bivar(mu1,sigma1,mu2,sigma2,a)};
\addplot3 [surf] {bivar(mu1,sigma1,mu2,sigma2,a)};
%\addplot3 [domain=-3:3,samples=31,patch refines=3, patch type=biquadratic, samples y=0, thick, smooth] (x,4,{normal(mu1,sigma1)});
%\addplot3 [domain=-3:3,samples=31, samples y=0, thick, smooth] (-1,x,{normal(mu2,sigma2)});

%$\draw [black!50] (axis cs:-1,0,0) -- (axis cs:4,0,0);
%$\draw [black!50] (axis cs:0,-1,0) -- (axis cs:0,4,0);

%\node at (axis cs:-1,1,0.18) [pin=165:$P(x_1)$] {};
%\node at (axis cs:1.5,4,0.32) [pin=-15:$P(x_2)$] {};
\end{axis}

\end{tikzpicture}} &\hspace*{-15pt}
 \resizebox{.3\textwidth}{!}{% \pgfplotsset{
% colormap={hot2}{color(0cm)=(orange!50!red!70!yellow!30); color(1cm)=(orange!50!red!70!yellow!90)}
% }
\pgfplotsset{
%  colormap={cool}{rgb255(0cm)=(220,220,220); rgb255(1cm)=(0,128,255); rgb255(2cm)=(255,0,255)}
  colormap={yg}{rgb255(0cm)=(232,240,232); rgb255(1cm)=(0,128,0)}
}

\begin{tikzpicture}[scale=0.75,
    declare function={mu1=0;},
    declare function={mu2=0;},
    declare function={a=1;},
    declare function={b(\a)=2/\a;},
    %declare function={gamma(\z)= 2.506628274631*sqrt(1/\z)+ 0.20888568*(1/\z)^(1.5)+ 0.00870357*(1/\z)^(2.5)- (174.2106599*(1/\z)^(3.5))/25920- (715.6423511*(1/\z)^(4.5))/1244160)*exp((-ln(1/\z)-1)*\z;},
    % Following from numerical recipies -- very accurate and stable
    declare function={gammaln(\z)= ln(2.506628274631 * (
    1.00000000019001 + 76.1800917294715/\z - 86.5053203294168/(\z + 1) + 24.0140982408309/(\z + 2) - 1.23173957245016/(\z + 3) + 1.20865097386618*0.001/(\z + 4) - 5.395239384953*0.000001/(\z + 5) 
    )) - \z - 4.5 + (\z - 0.5) * ln(\z + 4.5);},
    declare function={sigma1=0.5;},
    declare function={sigma2=1;},
    declare function={normal(\m,\s)=1/(2*\s*sqrt(pi))*exp(-(x-\m)^2/(2*\s^2));},
    declare function={bivar(\ma,\sa,\mb,\sb,\aa)=
exp ( (\aa-1) *  ln((x-\ma)^2/\sa^2 + (y-\mb)^2/\sb^2 + 0.001) - (\aa/2) *((x-\ma)^2/\sa^2 + (y-\mb)^2/\sb^2) - ln(pi*\sa*\sb) - gammaln(\aa) - \aa*ln(2) + \aa*ln(\aa) );}]
%* gamma(1)/gamma(\aa)/b(\aa)^(\aa) 
\begin{axis}[
    width=5.5cm,
    xlabel=$x_1$,
    ylabel=$x_2$,
    zlabel={$p(x_1,x_2)$},
    zmin=0,
    zmax=0.301,
    samples=24, %34
    domain=-3:3,
    y domain=-3:3,
    view={65}{65},
    %colormap name=whitered,
    %plot box ratio={1 1 5},
    %height=5cm,    
    %enlargelimits=false,
    %grid=major,
   %enlarge z limits
    %colorbar,
    %colorbar style={
    %    at={(1,0)},
    %    anchor=south west,
    %    height=0.25*\pgfkeysvalueof{/pgfplots/parent axis height},
    %    title={$P(x_1,x_2)$}
    %}
]
%\addplot3[surf,samples=30,shader=interp,domain=-3:3,colormap name=whitered] {bivar(mu1,sigma1,mu2,sigma2,a)}; 
%\addplot3[mesh,samples=7] {bivar(mu1,sigma1,mu2,sigma2,a)};
\addplot3 [surf] {bivar(mu1,sigma1,mu2,sigma2,a)};
%\addplot3 [domain=-3:3,samples=31,patch refines=3, patch type=biquadratic, samples y=0, thick, smooth] (x,4,{normal(mu1,sigma1)});
%\addplot3 [domain=-3:3,samples=31, samples y=0, thick, smooth] (-1,x,{normal(mu2,sigma2)});

%$\draw [black!50] (axis cs:-1,0,0) -- (axis cs:4,0,0);
%$\draw [black!50] (axis cs:0,-1,0) -- (axis cs:0,4,0);

%\node at (axis cs:-1,1,0.18) [pin=165:$P(x_1)$] {};
%\node at (axis cs:1.5,4,0.32) [pin=-15:$P(x_2)$] {};
\end{axis}

\end{tikzpicture}} & \hspace*{-15pt}
 \resizebox{.3\textwidth}{!}{% \pgfplotsset{
% colormap={gray}{color(0cm)=(orange!50!red!70!yellow!30); color(1cm)=(orange!50!red!70!yellow!90)}
% }

\pgfplotsset{
  %colormap={cool}{rgb255(0cm)=(220,220,220); rgb255(1cm)=(0,128,255); rgb255(2cm)=(255,0,255)}
  colormap={yg}{rgb255(0cm)=(232,240,232); rgb255(1cm)=(0,128,0)}
}

% \pgfplotsset{
%   colormap={greenyellow}{rgb255(0cm)=(0,128,0); rgb255(1cm)=(255,255,0)}
% }

\begin{tikzpicture}[scale=0.75,
    declare function={mu1=0;},
    declare function={mu2=0;},
    declare function={a=3;},
    declare function={b(\a)=2/\a;},
    %declare function={gamma(\z)= 2.506628274631*sqrt(1/\z)+ 0.20888568*(1/\z)^(1.5)+ 0.00870357*(1/\z)^(2.5)- (174.2106599*(1/\z)^(3.5))/25920- (715.6423511*(1/\z)^(4.5))/1244160)*exp((-ln(1/\z)-1)*\z;},
    % Following from numerical recipies -- very accurate and stable
    declare function={gammaln(\z)= ln(2.506628274631 * (
    1.00000000019001 + 76.1800917294715/\z - 86.5053203294168/(\z + 1) + 24.0140982408309/(\z + 2) - 1.23173957245016/(\z + 3) + 1.20865097386618*0.001/(\z + 4) - 5.395239384953*0.000001/(\z + 5) 
    )) - \z - 4.5 + (\z - 0.5) * ln(\z + 4.5);},
    declare function={sigma1=0.5;},
    declare function={sigma2=1;},
    declare function={normal(\m,\s)=1/(2*\s*sqrt(pi))*exp(-(x-\m)^2/(2*\s^2));},
    declare function={bivar(\ma,\sa,\mb,\sb,\aa)=
exp ( (\aa-1) *  ln((x-\ma)^2/\sa^2 + (y-\mb)^2/\sb^2 + 0.001) - (\aa/2) *((x-\ma)^2/\sa^2 + (y-\mb)^2/\sb^2) - ln(pi*\sa*\sb) - gammaln(\aa) - \aa*ln(2) + \aa*ln(\aa) );}]
%* gamma(1)/gamma(\aa)/b(\aa)^(\aa) 
\begin{axis}[
    width=5.5cm,
    xlabel=$x_1$,
    ylabel=$x_2$,
    zlabel={$p(x_1,x_2)$},
    zmin=0,
    zmax=0.301,
    %samples=33, %34
    %samples y=25,
    domain=-3:3,
    y domain=-3:3,
    view={65}{65},
    %colormap name=whitered,
    %plot box ratio={1 1 5},
    %height=5cm,    
    %enlargelimits=false,
    %grid=major,
   %enlarge z limits
    %colorbar,
    %colorbar style={
    %    at={(1,0)},
    %    anchor=south west,
    %    height=0.25*\pgfkeysvalueof{/pgfplots/parent axis height},
    %    title={$P(x_1,x_2)$}
    %}
]
%\addplot3[surf,samples=30,shader=interp,domain=-3:3,colormap name=whitered] {bivar(mu1,sigma1,mu2,sigma2,a)}; 
%\addplot3[mesh,samples=7] {bivar(mu1,sigma1,mu2,sigma2,a)};
\addplot3 [surf,mesh/ordering=y varies] table{datas.dat};
%\addplot3 [domain=-3:3,samples=31,patch refines=3, patch type=biquadratic, samples y=0, thick, smooth] (x,4,{normal(mu1,sigma1)});
%\addplot3 [domain=-3:3,samples=31, samples y=0, thick, smooth] (-1,x,{normal(mu2,sigma2)});

%$\draw [black!50] (axis cs:-1,0,0) -- (axis cs:4,0,0);
%$\draw [black!50] (axis cs:0,-1,0) -- (axis cs:0,4,0);

%\node at (axis cs:-1,1,0.18) [pin=165:$P(x_1)$] {};
%\node at (axis cs:1.5,4,0.32) [pin=-15:$P(x_2)$] {};
\end{axis}

\end{tikzpicture}}
  \end{tabular}
  \caption{\label{fig:one}\small EG density on $\reals^2$ with shape parameter $a=1/3$, $1$, and $3$ (from left to right). All displayed densities have equal covariances; the density corresponding to $a=q/2=1$ (middle) is a Gaussian density.}
\end{figure}

We focus on a particular non-Gaussian distribution: the \emph{Elliptical Gamma (EG) Distribution} (EGD)~\citep{kotz75,koutras86}. The (mean-zero) EG density (when it exists) for a point $\vx \in \reals^q$ is given by
\begin{equation}
  \label{eq:1}
  \peg(\vx ; \msigma, a, b) := \frac{\Gamma(q/2)}{\pi^{q/2} \Gamma(a) b^a|\msigma|^{1/2}} \bigl( \vx^\top  \msigma^{-1} \vx \bigr)^{a-q/2 } \exp \bigl ( -b^{-1}\vx^\top  \msigma^{-1} \vx\bigr),
  \end{equation}
where $\msigma \succ 0$ is the scatter matrix, and $a, b>0$ are scale and shape parameters~\citep{fang90}. Observe that~\eqref{eq:1} generalizes the Gaussian density (which corresponds to $a=q/2$) by reshaping it with an additional elliptical factor $(\vx^\top\msigma^{-1}\vx)^{a-q/2}$ that encodes different tail and peak behaviors---see Figure~\ref{fig:one} for an illustration. It is worth noting that for $a<q/2$ the EG density can be written as a scale mixture of Gaussians, using beta density as its scale (see Appendix~\ref{app:gsm}).

EGDs offer rich modeling power and are widely applicable: a mixture of mean-zero EGDs can approximate any symmetric distribution~\citep{fang90}. Moreover, EGDs are a subclass of \emph{Elliptically Contoured Distributions} (ECDs), which themselves are widely used in multivariate density estimation~\citep{ollila11}, Bayesian statistical data modeling~\citep{arellano06}, signal denoising~\citep{tan07}, financial data modeling~\citep{bingham02}, pattern recognition~\citep{theiler10}, and many other applications. Likewise, mixtures of ECDs have also found widespread use, e.g., in robust statistical modeling~\citep{lange89}, denoising~\citep{rabbani08}, signal processing, among others---the survey \citep{ollila11} provides several more applications and references.

A further motivation for our work is its potential for enabling robust recovery of multiple subspaces~\citep{lerman2011}, where the ``robustness'' refers to being able to estimate the subspace even when only a certain percentage of data lie in the subspace. This topic in turn has various applications in  unsupervised learning, computer vision, and biomedical engineering---see e.g.,~\citep{soltanolkotabi12}.

We note that EGDs have an unbounded influence function, so maximum likelihood estimates are not robust in the usual sense. That is, in the presence of outliers, the EGD scatter matrix estimates will get skewed. But this sensitivity is not restrictive in practice: if the subspaces have outliers, their impact on the ML estimates  can be countered by using a mixture model that contains a non-informative uniform distribution as an additional component~\citep{hennig2004breakdown}.

\subsection{Summary of main results}
We study the following two interrelated tasks for EGDs: (i) maximum likelihood (ML) estimation; and (ii) parameter estimation for a mixture model. Task (i) presents the main theoretical challenges. Its associated maximization problem may be nonconcave, and moreover, efficiently imposing the constraint $\msigma \succ 0 $ is nontrivial.

More specifically, we develop ML estimation procedures for concave ($a \ge q/2$) as well as nonconcave ($a < q/2$) EGD log-likelihoods. Our procedures are cast as \emph{non-Euclidean} fixed-point algorithms, each of whose two cases has a  rather different convergence analysis. We first tackle the concave case, where the key difficultly lies in efficiently handling the positive-definiteness constraint. Next we handle the harder nonconcave case, where not only must we fulfill positive-definiteness but also obtain global optimality despite nonconcavity. Finally, we use our ML algorithms as subroutines of a modified EM algorithm applied to EGD mixture models.

We experiment with both simulated and real data and observe large speedups over state-of-the-art manifold optimization algorithms as well as over a well-known iteration of \citet{kent91} (which incidentally applies only to a subset of the cases amenable to our methods). Implementations of our methods can be found in our larger software package on mixture modeling~\citep{hosseini2015mixest}.

\section{Background}
EG distributions are subclass of ECDs. A $q$-dimensional random vector $\mx$ is distributed according to an ECD with \emph{mean} $\vm \in \mathbb{R}^q$ and \emph{scatter} $\msigma \in \mathbb{R}^{q\times q}$,
if its characteristic function is of the form 
$\Phi_X(\vt) = \exp(i \, \vt^\top \vm)g(\vt^\top \msigma \vt)$, for some function $g : \reals_+ \to \reals$. If it exists, the density of an ECD assumes the form
\begin{equation*}
p_X(\vx)= |\msigma|^{-1/2} f \bigl((\vx-\vm)^\top  \msigma^{-1} (\vx-\vm)\bigr),
\end{equation*}
for a suitable function $f : \reals_+ \to \reals$. We focus on mean-zero ECDs, so that
\begin{equation}
\label{eqn_1a}
p_X(\vx) = |\msigma|^{-1/2} f \bigl(\vx^\top  \msigma^{-1} \vx\bigr).
\end{equation}
Therewith $\mx$ factors into a uniform hypspherical component and a scaled-radial part, i.e., $\mx=R\msigma^{1/2}\mU$ with $\mU \sim \text{Unif}(\mathbb{S}^{q-1})$ and $R$ a univariate random variable given by $R=\enorm{\msigma^{-1/2}\mx}$~\citep{fang90B}. The random variable $R$ has the density:
\begin{equation*}
  p_R(r) := 2\pi^{q/2} f(r^2)r^{q-1}/\Gamma(\tfrac q 2).
\end{equation*}
Thus, the square radial component $\Upsilon=R^2$ has the density $p_{\Upsilon}(\upsilon) := \pi^{q/2} f(\upsilon)\upsilon^{q/2-1}/\Gamma(\tfrac q 2)$. When this square radial component is distributed according to a gamma distribution we obtain an EGD. Recall that a gamma-distributed random variable has density
\begin{equation}
\label{eqn_2a}
p_{\text{ga}}(\upsilon;a,b) = \upsilon^{a-1}\Gamma(a)^{-1}b^{-a}\exp\left(-\upsilon/b\right),
\end{equation}
where $a$ is the \emph{shape} parameter and $b$ is a \emph{scale} parameter. Using~\eqref{eqn_2a} as the radial density, we obtain the density generating function $f$ for \eqref{eqn_1a}, which then yields the EGD density~\eqref{eq:1}. If $\msigma$ equals the distribution covariance, i.e., $\msigma=\E[\mx \mx^{\top}]$, then $b=q/a$ (see~\citealt[Eq.~2.16]{fang90B}).

\subsection{ML estimation}
Obtaining closed-form ML estimates for ECDs is typically impossible, though in special cases such as multivariate $t$-distributions, a recursive algorithm is known~\cite{lange89}. For a wider review of ML estimation for ECDs see \cite{ollila11} and references therein; see also~\citep{sra15,zhang13}. 

A well-known fixed-point algorithm for estimating the scatter matrix of ECDs is due to  \citet{kent91}. Their algorithm is applicable for a general class of ECDs, including the nonconcave case ($a < q/2$) of EGDs. \citet{duembgen2015new} propose a generic method for improving the convergence speed of Kent and Tyler's iterations. \citet{sra15} propose different fixed-point algorithms applicable to a broad class of ECDs. We propose below a new algorithm similar to that  in~\citep{sra15}, but with a different convergence analysis specialized to EGDs. A notable property of our convergence analysis is its lack of dependence of any existence result, a prerequisite of all previous results. Therefore, it applies even to the cases where the solution is a singular matrix. 

A further interesting aspect is that for the nonconcave case, an EGD can be expressed as a scale mixture of Gaussians (Appendix~\ref{app:gsm}). The algorithm due to \citet{kent91} applied to EGDs can be viewed as a majorization-maximization method \citep{sra15}, or also as an EM algorithm for estimating parameters of the scale mixtures of Gaussians. There is a broad literature for accelerating EM algorithm for scale mixtures of Gaussians. For example, \citet{meng1997algorithm} proposed an algorithm for accelerating EM algorithm for multivariate $t$-distributions.

\section{Maximum likelihood parameter estimation}
\label{sec:mle}
In this section we derive new ML estimation procedures for EGDs covering both concave and nonconcave log-likelihoods. Let $\{\vx_i\}_{i=1}^n$ be a set of i.i.d.\ samples from a mean-zero EGD with unknown scatter $\msigma$. The log-likelihood of the samples $\{\vx_i\}_{i=1}^n$ is
\begin{equation}
\label{eq.ll}
\begin{split}
\ell(a,b,\msigma) :=
\text{const.}-\tfrac{n}{2}\log |\msigma| +\nlsum_{i=1}^{n}\bigl [\bigl (a-\tfrac q 2 \bigr)\log  (\vx_i^\top \msigma^{-1}\vx_i)-\tfrac{1}{b}\vx_i^\top \msigma^{-1}\vx_i \bigr].
\end{split}
\end{equation}
We estimate $\msigma$ assuming $a$ and $b$ are given. This task splits into two natural cases:
\begin{enumerate}[(i)]
  \setlength{\itemsep}{0pt}
\item \emph{Concave.} Here $a \ge q/2$ and $\ell$ is concave in $\msigma^{-1}$ (though not in $\msigma$).
\item \emph{Nonconcave.} Here $a < q/2$, so the second term in~\eqref{eq.ll} is no longer concave.\end{enumerate}

Clearly, if $\ell$ is strictly concave and attains its maximum, this must be unique. More remarkably, even when $\ell$ is nonconcave, we will see that its hidden geometric structure ensures uniqueness (shown at the end of this section).
We note that the content of this section up to \eqref{eq:6b} also follows from more general results on ECDs~\citep{maronna76}. We present the details to set notation and for making our exposition self-contained.

Since the constraint $\msigma \succ 0$ is an open set, we can use the gradient based necessary condition $\nabla_{\msigma}\ell=0$. Moreover, since~\eqref{eq.ll} has a unique global maximum, a positive definite solution to $\nabla_{\msigma}\ell=0$ must be the desired ML estimate. Consider therefore the following nonlinear equation obtained upon differentiating $\ell$:
\begin{equation}
\label{eq:2}
-\tfrac{n}{2}\msigma^{-1}-\bigl(a-\tfrac q 2\bigr) \nlsum_{i=1}^{n}{\frac{\msigma^{-1}\vx_i\vx_i^\top \msigma^{-1}}{\vx_i^\top \msigma^{-1}\vx_i}}+\tfrac{1}{b}\nlsum_{i=1}^{n}{\msigma^{-1}\vx_i\vx_i^\top \msigma^{-1}}=0.
\end{equation}
Now add $\frac n 2 \msigma^{-1}$ to both sides and rescale by $\sqrt{\tfrac2n}\msigma^{1/2}$ to obtain the equation
\begin{equation}
  \label{eq:6}
M(\msigma,c,d) :=
c\nlsum_{i=1}^{n}{\frac{\msigma^{-1/2}\vx_i\vx_i^\top \msigma^{-1/2}}{\vx_i^\top \msigma^{-1}\vx_i}} +d\nlsum_{i=1}^{n}{\msigma^{-1/2}\vx_i\vx_i^\top \msigma^{-1/2}}=\mi;
\end{equation}
where we have introduced the constants
\begin{equation}
  \label{eq:7}
  c := -\frac{2\left (a-q/2 \right)}{n},\qquad d :=\frac{2}{bn}.
\end{equation}

We now state our uniqueness theorem, which shows that upon its existence, the solution to~\eqref{eq:6} is unique.
\begin{theorem}
  \label{thm.uniqml}
  If the data set $\left \{\vx_i \right \}_{i=1}^{n}$ spans $\reals^{q}$ and $\msigma_1$, $\msigma_2$ are positive definite matrices for which $\mm(\msigma_1,c,d)=\mm(\msigma_2,c,d)$ and $c>0$, then $\msigma_1=\msigma_2$.
\end{theorem}
\begin{proof}
  See Appendix~\ref{app:uniqueML}.\footnote{An alternative proof also follows from~\citep[Thm.~2.2]{kent91}.}
\end{proof}

To solve~\eqref{eq:6}, we present two fixed-point algorithms depending on the sign of $c$. We rewrite~\eqref{eq:6} in a form more amenable to analysis. First, we introduce a matrix $\bm B$ and transformed vectors $\vy_i$ ($i=1,2,\ldots,n$) defined as
\begin{equation}
\label{eqn.B}
\mb=d\nlsum_{i=1}^{n}{\vx_i\vx_i^\top},\quad \vy_i=\mb^{-1/2}\vx_i.
\end{equation}
Then, set $\mgamma=\mb^{-1/2}\msigma\mb^{-1/2}$ and note that $\mgamma^{1/2}$ has the form $\mb^{-1/2}\msigma^{1/2}\mq^{\top}$ for some orthogonal matrix $\mq$. This observation allows us to rewrite~\eqref{eq:6} as
\begin{equation}
  \label{eq:6b}
c\sum_{i=1}^{n}\frac{\mgamma^{-1/2}\vy_i\vy_i^\top \mgamma^{-1/2}}{\vy_i^\top 
\mgamma^{-1}\vy_i}+\mgamma^{-1}=\mi.
\end{equation}
From a solution $\mgamma^*$ to~\eqref{eq:6b}, we recover $\msigma^* =\mb^{1/2}\mgamma^*\mb^{1/2}$ as the solution to~\eqref{eq:6}. 

Our algorithms for solving~\eqref{eq:6b} split into two cases: (1) concave ($a \ge q/2$, equivalently $c\leq0$); and (2) nonconcave ($a < q/2$, equivalently $c > 0$). 

\subsection{The concave case: $c \leq 0$}
We omit $c=0$ as it is trivial. Rearrange~\eqref{eq:6b} and consider the following ``positivity-preserving'' iteration
\begin{equation}
\label{eq:3}
\mgamma_{p+1}= \biggl (-c\sum_{i=1}^{n}{\frac{\mgamma_{p}^{-1/2}\vy_i\vy_i^\top \mgamma_{p}^{-1/2} }{\vy_i^\top  \mgamma_{p}^{-1}\vy_i}}+\mi \biggr)^{-1},\qquad p=0,1,\ldots,
\end{equation}
which by construction ensures that if $\mgamma_p \succ 0$, then $\mgamma_{p+1} \succ 0$. Clearly, any limit point of iteration~\eqref{eq:3} is also positive definite and satisfies \eqref{eq:6b}. This suffices for global optimality as the log-likelihood is strictly concave for $c<0$. Any limit point of iteration~\eqref{eq:3} is a fixed point of the following map (where $\ms = \mgamma^{-1} \succ 0$):
\begin{equation}
    \label{eq.3}
    \Gc \equiv \ms \mapsto \mi + c^\prime \ms^{1/2}\my\md_{\ms}\my^\top \ms^{1/2},
  \end{equation}
where $\my$ is a matrix with $\vy_i$ as its $i$th column, $\md_{\ms} = \Diag(1/\vy_i^\top\ms\vy_i)$, and $c^\prime=-c$. Therefore, we prove convergence of iteration~\eqref{eq:3} by showing the map $\Gc$ to be a fixed-point map. This is proved by Theorem~\ref{thm.one}, the main result of this section. 
\begin{theorem}
  \label{thm.one}
Let $\ms_0\in D$ (for the compact set $\Dc$ defined by Lemma~\ref{lem.two}) be chosen arbitrarily. Then, the iteration $\ms_{k+1} = \Gc(\ms_k)$ converges to a unique fixed-point $\ms^*$.
\end{theorem}

Our proof relies on the following crucial lemma which establishes existence of a compact set $\Dc$, within which the map $\Gc$ has a fixed-point. 
\begin{lemma}
  \label{lem.two}
  Let $\Gc$ be as in~\eqref{eq.3}; let $\Dc := [\mi, \mu \mi]$, where $\mu > (1+c^\prime n)$, then $\Gc(\Dc) \subset \Dc$.
\end{lemma}
\begin{proof}
Let $\Dc = [\mi, \mu \mi]$, where $\mu > 1$ is some scalar (to be determined). We show that there exists a $\mu$ such that $\Gc(\Dc) \subset \Dc$, i.e., if $\ms \in \Dc$, then $\Gc(\ms) \in \text{int}(\Dc)$.

First, check that if $\ms=\mi$, then $\Gc(\ms) = \mi + c^{\prime}  \my D_{\mi}\my^\top \prec \mu \mi$ for suitably large $\mu$. Moreover, $\Gc(\mi) \succ \,\mi$ (since $c^\prime  > 0$ and $\my \neq 0$). If $\ms = \mu \mi$, then $\Gc(\ms) = \mi + c^\prime  \my \md_{\mi}\my^\top$ (the $\mu$ cancels out because $\Gc(\alpha \ms)=\Gc(\ms)$ for all $\alpha > 0$). Thus, if $\ms=\mi$ or $\ms=\mu\mi$, then $\mi \prec \Gc(\ms) \prec \mu \mi$. 

It remains to show that if $\mi \prec \ms \prec \mu \mi$, then $\mi \prec \Gc(\ms) \prec \mu \mi$. Notice, however that $\frac{\ms^{1/2}\vy_i\vy_i^\top\ms^{1/2}}{\vy_i^\top\ms\vy_i} = \frac{\vz_i\vz_i^T}{\vz_i^\top\vz_i} \preceq \mi$. Thus, $\ms^{1/2}\my\md_{\ms}\my^\top\ms^{1/2} \preceq n\mi$, so that we have the inequality
\begin{equation}
  \label{eq.4}
  \mi \prec \Gc(\ms) \preceq \mi + c^\prime  n \mi= (1+c^\prime  n) \mi.
\end{equation}
Therefore, if $\mu > (1+c^\prime  n)$, we are guaranteed that $\Gc(\ms) \prec \mu \mi$. 
\end{proof}

\begin{corr}
  \label{cor.exist}
  The map $\Gc$ defined by~\eqref{eq.3} has a fixed point in $\Dc$.
\end{corr}
\begin{proof}
Lemma~\ref{lem.two} shows that $\Gc$ maps a compact convex set $\Dc$ to itself, so that using continuity of $\Gc$ on $\Dc$ and Brouwer's fixed-point theorem~\citep{granas03}, we conclude that $\Gc$ has a fixed point in $\Dc$.  
\end{proof}
Corollary~\ref{cor.exist} proves existence, while Theorem~\ref{thm.one} proves the \emph{harder} result that this fixed point is \emph{unique}. And more importantly, that this fixed-point can be computed by simply iterating $\Gc$ (Picard iteration). Before proving this claim, we need one more step.

\begin{prop} 
  \label{prop.iterated.map}
  Let $\Gc$ be a continuous map that maps a nonempty set $\Dc$ to itself. If the iterated map $\Gc^m$ has a unique fixed point for every integer $m\ge 1$, then beginning with $\ms_0 \in \Dc$ the Picard iteration $\ms_{k+1}=\Gc(\ms_k)$ converges to this unique fixed point.
\end{prop}
\begin{proof}
  Classic result in fixed-point theory; see e.g.,~\citep{khamsi}. 
  \end{proof}

\noindent Using Proposition~\ref{prop.iterated.map}, we are now ready to prove Theorem~\ref{thm.one}. 

\begin{proof}[Proof of Theorem~\ref{thm.one}.]
  As mentioned above, Brouwer's theorem shows that there exists a fixed point of the map $\Gc$. But this assertion does not imply that the iteration $\ms_{k+1}=\Gc(\ms_k)$ converges to this (or any) fixed point. However, for $c < 0$, we know that the  log-likelihood is strictly concave, whereby, if it attains its maximum, this maximum must be unique. Any fixed-point of the map $\Gc$ satisfies the first-order necessary and sufficient (due to concavity) conditions, so that there can be only a single unique fixed-point. But before we conclude that the iteration $\ms_{k+1}=\Gc(\ms_k)$ takes us to this unique fixed point by merely iterating $\Gc$, we need an additional argument.

  The key idea is to apply Brouwer repeatedly. First, observe that for each $m \ge 1$, the map $\Gc^m$ has a unique fixed-point: this is so, because inductively one can verify that $\Gc^m(D) \subset D$, and since $D$ is compact, Brouwer's theorem along with concavity of the log-likelihood lets us conclude that $\Gc^m$ has a unique fixed point. Now we appeal to Proposition~\ref{prop.iterated.map}, which shows that iterating $\ms_{k+1}=\Gc(\ms_k)$ yields the fixed-point.
\end{proof}

\subsection{The nonconcave case: $c > 0$}
Iteration~\eqref{eq:3} does not apply to $c>0$ since positive definiteness of the iterates can no longer be guaranteed.  Therefore, we rewrite~\eqref{eq:6b} differently. Multiplying it on the left and right by $\mgamma^{1/2}$ and introducing a new parameter $\alpha>0$, we arrive at the iteration
\begin{equation}
\label{eqn_iter}
\mgamma_{p+1} = \alpha_p \mgamma_p^{1/2}\mn_p\mgamma_p^{1/2},\qquad p=0,1,\ldots,
\end{equation}
where $\alpha_p>0$ is a free scalar parameter, and the matrix $\mn_p$ is given by
\begin{equation}
  \label{eqn_2i2}
 \mn_p= c\sum_{i=1}^{n}{\frac{\mgamma_p^{-1/2}\vy_i\vy_i^\top \mgamma_p^{-1/2} }{\vy_i^\top  \mgamma_{p}^{-1}\vy_i}}+\mgamma_p^{-1}.
\end{equation}
We show that under a specific choice of the sequence $\{\alpha_p\}$, iteration~\eqref{eqn_iter} converges, and that in addition $\alpha_p \to \alpha^*=1$. Thus, $\lim_{p\to\infty}\mgamma_{p}=\mgamma^*$ satisfies \eqref{eqn_iter}, whereby $\mgamma^*$  is the desired ML solution. 

Our proof relies on a key technical result (Lemma~\ref{lem:one}), which shows that one can find $\alpha_p$ values that lead to an increase in the smallest eigenvalue of  $\mn_p$ and a decrease of the largest eigenvalue of $\mn_p$.

\begin{lemma}
\label{lem:one}
Let $\lambda_{1,p}>\alpha_p^{-1}$ and $\lambda_{q,p}<\alpha_p^{-1}$ represent the largest and smallest eigenvalues of $\mn_p$, respectively. If the data set $\left \{\vy_i \right \}_{i=1}^{n}$ spans $\mathbb{R}^{q}$ then $\lambda_{1,p+1}\leq\lambda_{1,p}$ and $\lambda_{q,p+1}\geq\lambda_{q,p}$.
\end{lemma}
\begin{proof}
  See Appendix~\ref{sec:lemone}.
\end{proof}
The main result of this section is Theorem~\ref{thm:two}, which shows that there is a sequence $\set{\alpha_p}\to 1$, for which~\eqref{eqn_2i2} converges.
\begin{theorem}
\label{thm:two}
Let $\lambda_{1,p}\geq 1$ and $\lambda_{q,p}\leq 1$ represent the largest and smallest eigenvalues of $\mn_p$ respectively. If the data set $\left \{\vx_i \right \}_{i=1}^{n}$ spans $\mathbb{R}^{q}$, then one can find an $\alpha_p$ such that $1 \leq \lambda_{1,p+1}\leq\lambda_{1,p}$ and $\lambda_{q,p}\leq\lambda_{q,p+1}\leq 1$. Thus,  $\alpha_p \to 1$ and iteration~\eqref{eqn_2i2} converges. 
\end{theorem}
\begin{proof}
Define $\mgamma' = \mgamma_p^{1/2}\mn_p\mgamma_p^{1/2}$  and
\begin{align*}
\mn'= c\sum_{i=1}^{n}{\frac{\mgamma'^{-1/2}\vy_i\vy_i^\top \mgamma'^{-1/2} }{\vy_i^\top  \mgamma'^{-1/2}\vy_i}}+\mgamma'^{-1}.
\end{align*}
Rewriting $\mn_{p+1}$ in terms of $\mgamma'$ and $\mn'$, we obtain $\mn_{p+1} =\mn' +(\alpha_p^{-1} - 1) \mgamma'$.
Now we consider three cases for the eigenvalues of $\mn'$ and express the possible values of $\alpha_p$.

\noindent\textbf{1).}\hskip6pt $\lambda_1' \geq 1$ and $\lambda_q'\leq 1$: In this case $\alpha_p=1$ is the solution.

\noindent\textbf{2).}\hskip6pt $\lambda_1'\leq 1$ and $\lambda_q'\leq 1$: If $\alpha_p$ decreased toward zero, then $\lambda_{1,p+1}$ increases toward infinity. In addition $\lambda_{1,p+1}$ is a continuous function of $\alpha_p$, therefore if we increase $\alpha_p$ it will be possible to find an $\alpha_p$ such that $\lambda_{1,p+1}=1$. At the same time, because of the previous lemma, since $\lambda_{q,p}\leq 1 \leq \alpha_p^{-1}$, we have $\lambda_{q,p+1}\geq\lambda_{q,p}$. Note that finding $\alpha_p$ is equivalent to an eigenvalue problem: we want to find an $\alpha_p$, such that the largest eigenvalue of $\mn_{p+1}$ becomes one. One can show that $\alpha_p^{-1}$ should be the smallest eigenvalue of the following matrix (see Case I at the end of the proof for the derivation):  
\begin{eqnarray}
\mgamma'-c\sum_{i=1}^{n}{\frac{\vy_i\vy_i^\top }{\vy_i^\top 
\mgamma'^{-1}\vy_i}},
\label{eqn_IIId}
\end{eqnarray}
\noindent\textbf{3).}\hskip6pt $\lambda_1'\geq 1$ and $\lambda_q' \geq 1$: If $\alpha_p$ increases toward infinity, then $\lambda_{q,p+1}\longrightarrow \lambda<1$. Similar to the previous case $\lambda_{q,p+1}$ is a continuous function of $\alpha_p$, therefore if we decrease $\alpha_p$, it will be possible to find an $\alpha_p$ such that $\lambda_{q,p+1}=1$. At the same time since $\lambda_{1,p}\geq 1\geq \alpha_p^{-1}$, we have $\lambda_{1,p+1}\leq\lambda_{1,p}$. One can show that in this case $\alpha_p^{-1}$ should be the largest eigenvalue of the matrix given in~\eqref{eqn_IIId} (see Case II at the end of the proof).

Since the sequences $\lambda_{1,p}$ and $\lambda_{q,p}$ are both bounded and they are decreasing and increasing respectively, they are convergent. From the explained procedure for finding $\alpha_{p}$, it is easy to see that the convergent value of $\lambda_{1,p}$ and $\lambda_{q,p}$ satisfy the first case, and therefore $\alpha_p$ converges to one. We present the remaining details below.

\noindent\textbf{Case I:}

We want to calculate $\alpha_p^{-1}$ such that the largest eigenvalue of matrix $\mn_{p+1}$ below becomes one, given that the eigenvalues corresponding to $\alpha_p=1$ are smaller than one:
\begin{equation}
\mn_{p+1}=c\sum_{i=1}^{n}{\frac{\mgamma'^{-1/2}\vy_i\vy_i^\top \mgamma'^{-1/2}}{\vy_i^\top 
\mgamma'^{-1}\vy_i}}+\alpha_p^{-1}\mgamma'^{-1}.
\label{eqn_IIIa}
\end{equation}
Since both matrices $\mgamma'^{-1}$ and $c\sum_{i=1}^{n}{\frac{\mgamma'^{-1/2}\vy_i\vy_i^\top \mgamma'^{-1/2}}{\vy_i^\top 
\mgamma'^{-1}\vy_i}}$ are positive definite, the largest eigenvalue increases if we increase $\alpha_p^{-1}$. Therefore, we need to find the smallest $\alpha_p^{-1}$ such that an eigenvalue of the matrix $\mn_{p+1}$ becomes one. We have the eigenvalue problem
\begin{eqnarray}
\mn_{p+1}\vu=\vu.
\label{eqn_IIIb}
\end{eqnarray}
Assume $\vu=\mgamma'^{1/2}\vv$ and multiply~\eqref{eqn_IIIb} from left by $\mgamma'^{1/2}$ to  obtain
\begin{eqnarray}
c\sum_{i=1}^{n}{\frac{\vy_i\vy_i^\top }{\vy_i^\top 
\mgamma'^{-1}\vy_i}}\vv+\alpha_p^{-1}\vv=\mgamma'\vv.
\end{eqnarray}
Rearranging this equation we then obtain
\begin{eqnarray}
\mgamma'\vv-c\sum_{i=1}^{n}{\frac{\vy_i\vy_i^\top }{\vy_i^\top 
\mgamma'^{-1}\vy_i}}\vv=\alpha_p^{-1}\vv.
\label{eqn_IIIc}
\end{eqnarray}
Hence, $\alpha_p^{-1}$ is the smallest eigenvalue of the the matrix in \eqref{eqn_IIId}.

\noindent\textbf{Case II:}

We want to calculate $\alpha_p^{-1}$ such that the smallest eigenvalue of the matrix in \eqref{eqn_IIIa} becomes one given that eigenvalues of the matrix for $\alpha_p=1$ are larger than one. Since both matrices $\mgamma'^{-1}$ and $c\sum_{i=1}^{n}{\frac{\mgamma'^{-1/2}\vy_i\vy_i^\top \mgamma'^{-1/2}}{\vy_i^\top 
\mgamma'^{-1}\vy_i}}$ are positive definite, if we decrease $\alpha_p^{-1}$, the smallest eigenvalue decreases. Therefore, we need to find the largest $\alpha_p^{-1}$ such that an eigenvalue of the matrix $\mn_{p+1}$ becomes one. Here, we have the same eigenvalue problem, so we obtain that $\alpha_p^{-1}$ is the largest eigenvalue of the matrix in \eqref{eqn_IIId}.
\end{proof}

\begin{rmk} We could have invoked a result of~\citep{kent91}, or the more general theory of~\citep{sra13} to obtain convergence proofs of a different fixed-point iteration that computes $\mgamma$. However, the convergence results of~\citep{kent91,sra13} depend on the existence of an ML solution. 

In contrast, Theorem~\ref{thm:two} proves a \textbf{stronger} result because it does not depend on any existence requirement on the ML solution. This generality has some important consequences: if the ML solution exists, then inevitably iteration~\eqref{eqn_2i2} converges to it. But when the ML solution does not exist (which is well possible), then the iterative algorithm still converges, though now the convergent solution is singular. This singular matrix possesses specific structure that can be then used for robust subspace recovery, generalizing the subspace recovery approach of~\citep{zhang2015robust}. 

Furthermore, Theorem~\ref{thm:two} yields a computationally more efficient method that outperforms not only the methods of~\citep{kent91} and \citep{sra13} but also sophisticated manifold optimization techniques (see Section~\ref{sec:expt}).
\end{rmk}    

\begin{rmk}
\label{rmk.1}
  The above theorem suggests $\alpha_p$ values which are not necessarily optimal, though easy to calculate. In practice, we observed that choosing $\alpha_p$ such that the trace of the matrix $\mn_{p+1}$ becomes $q$, that is $\alpha_p=\text{tr}(\mgamma'^{-1})/(2a)$, leads to faster convergence for smaller values of $a$. However, for this case our convergence proof does not apply. 
When $b=q/a$ and $a \to 0$, then it is easy to see that with this choice of $\alpha_p$, the proposed fixed point algorithm actually converges to the M-estimator of scatter matrix~\citep{Tyler1987}. This M-estimator is equal to the ML estimate of an angular central Gaussian distribution~\citep{tyler1987b}. It is consistent with a recent result showing that KL-divergence between EG distributions converges to the KL-divergence between angular central Gaussian distributions, when  $a \to 0$~\citep{zadeh2015}.
\end{rmk}

\section{Experimental results for ML estimation}
\label{sec:expt}
We report results on the convergence speed of our fixed-point iterations~\eqref{eq:3} and~\eqref{eqn_2i2}. We compare our algorithms against three (Riemannian) manifold optimization techniques, namely a trust-region method, the conjugate gradient method and limited-memory BFGS (LBFGS) method, and against the Kent-Tyler iteration~\citep{kent91}. We used the Manopt toolbox for manifold optimization~\citep{manopt}, except for LBFGS, which we implemented ourselves~\citep{sra15}. We also tested other optimization techniques such as semidefinite programming (SDP) solvers based on interior-point methods~\citep{nestNem94} (the convex case); Table~\ref{tab:sdp} reports representative running time results. The SDP solvers run much slower than methods adapted to the manifold, so for our other experiments we limit our attention to manifold optimization and fixed-point methods. 
\begin{table}[htbp]
  \begin{tabular}{l|l|c}
    Solver         & Time & Negative log-likelihood\\
    \hline
    SDPT3       & 93s            & $1.4245 \times 10^4$\\
    SeDUMI      & 47s            & $1.4245 \times 10^4$\\
    Manopt CG   & 0.50s          & $1.4245 \times 10^4$\\ 
    Fixed-Point &\textbf{0.15s}  & $1.4245 \times 10^4$\\
    \hline
  \end{tabular}
  \caption{Speed comparison between our fixed point iteration, manifold optimization, and two standard convex programming solvers. The dimensionality of the data is $d=8$, the shape parameter of the EG density is $a=20$, and number of observations is $n=1000$. As shown in the third column, all methods attain the same negative log-likelihoods.}
  \label{tab:sdp}
\end{table}

\begin{figure}
\resizebox{.45\textwidth}{!}{\input{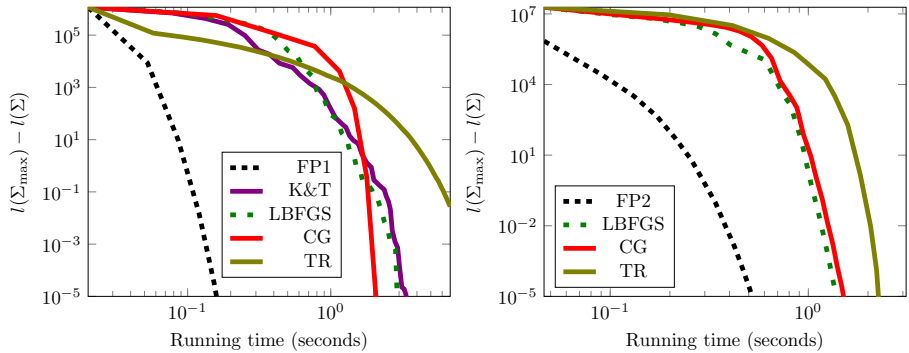}}
\resizebox{.45\textwidth}{!}{% This file was created by matlab2tikz v0.4.7 (commit 3442858e5a642c135c5e9dab6a960bee5b9c6f8d) running on MATLAB 7.8.
% Copyright (c) 2008--2014, Nico Schlömer <nico.schloemer@gmail.com>
% All rights reserved.
% Minimal pgfplots version: 1.3
% 
% The latest updates can be retrieved from
%   http://www.mathworks.com/matlabcentral/fileexchange/22022-matlab2tikz
% where you can also make suggestions and rate matlab2tikz.
% 
\begin{tikzpicture}

\begin{axis}[%
scale=0.6,
width=4.38666666666667in,
height=3.50555555555556in,
scale only axis,
xmode=log,
xmin=0.0463404505494506,
xmax=3.114550386948,
xminorticks=true,
xlabel={Running time (seconds)},
ymode=log,
ymin=9.99000999000999e-06,
ymax=19130613.5734402,
yminorticks=true,
legend pos=south west,
ylabel={$l(\Sigma_{\max})-l(\Sigma)$}
]
\addplot [color=black,dashed,line width=2.5pt]
  table[row sep=crcr]{0.046386791	701917.53159077\\
0.09035406	26506.2485945773\\
0.130801482	3347.24326002528\\
0.173222437	418.243928756216\\
0.2132196	51.5651687739883\\
0.256598677	6.32226299314061\\
0.295472079	0.773954199277796\\
0.339716869	0.0947459205635823\\
0.378020355	0.0116051288205199\\
0.421990924	0.00142254476668313\\
0.463004091	0.000174516753759235\\
0.505410164	2.14297324419022e-05\\
0.545370658	2.6369234547019e-06\\
0.588464502	3.24624124914408e-07\\
0.62802075	4.37721610069275e-08\\
0.671660359	8.49831849336624e-09\\
0.712240284	8.73114913702011e-10\\
0.758217943	2.27009877562523e-09\\
};
\addlegendentry{FP2};

\addplot [color=black!50!green,loosely dashed,line width=2.5pt]
  table[row sep=crcr]{0.046386791	19111502.0713688\\
0.103024846	9137182.63679444\\
0.161109326	6477533.53336068\\
0.219241985	4351134.33917834\\
0.279558342	2857355.05136926\\
0.341673056	1406929.83678065\\
0.404929154	472318.17528551\\
0.604647925	109750.88478885\\
0.679780179	16965.3206179521\\
0.744215415	3816.81512229616\\
0.810552028	1421.39238605002\\
0.922983271	40.8582281257841\\
1.033456041	0.572795458720066\\
1.14319876	0.0198791624861769\\
1.255538447	0.000556702318135649\\
1.371203265	1.40880001708865e-05\\
1.491319429	5.4424162954092e-07\\
1.610531431	1.58906914293766e-08\\
1.730005041	3.89991328120232e-09\\
1.870638292	4.65661287307739e-10\\
1.970595089	4.07453626394272e-10\\
2.109811305	2.3283064365387e-10\\
2.253952365	0\\
3.111438948	0\\
};
\addlegendentry{LBFGS};

\addplot [color=red,solid,line width=2.5pt]
  table[row sep=crcr]{0.046386791	19111502.0713688\\
0.110240384	9137182.63679444\\
0.182429461	6077066.80667506\\
0.249465986	4521790.02378379\\
0.31707805	3530264.18544667\\
0.383430094	2692093.96538979\\
0.448627389	1854972.66878935\\
0.516431249	1066588.11432355\\
0.583153615	466476.404646119\\
0.652221397	130567.72391673\\
0.720753902	14001.0370275455\\
0.806891901	3354.04738489434\\
0.875341299	1037.52693146345\\
0.944271299	75.7723747254349\\
1.029134963	9.1461083451868\\
1.09660914	1.08287826477317\\
1.177916833	0.126975666964427\\
1.245896333	0.0148469270789064\\
1.327953565	0.00173521513352171\\
1.395724749	0.000202872906811535\\
1.478304877	2.37331842072308e-05\\
1.544352196	2.77895014733076e-06\\
1.625359405	3.25730070471764e-07\\
1.692609735	4.15602698922157e-08\\
1.830276859	5.99538907408714e-09\\
1.957110396	1.97906047105789e-09\\
2.153207324	1.97906047105789e-09\\
};
\addlegendentry{CG};

\addplot [color=red!50!green,solid,line width=2.5pt]
  table[row sep=crcr]{0.046386791	19111502.0713688\\
0.197439273	9156687.66757022\\
0.414337104	3206286.84007033\\
0.631099593	905238.324376011\\
0.849690427	229303.980607772\\
0.990673161	77920.2321737307\\
1.216624734	17268.0526432164\\
1.362188494	3107.23140349583\\
1.581242008	191.688102479326\\
1.828440757	1.4805232239305\\
2.062302448	0.00932682416168973\\
2.210565069	0.000129466003272682\\
2.429173682	1.39698386192322e-09\\
2.653733243	2.73576006293297e-09\\
3.036229856	2.27009877562523e-09\\
};
\addlegendentry{TR};
%\node[fill=white, inner sep=0mm, rotate=300.347856981794, font=\bfseries, text=black!50!blue, draw=black]
at (axis cs:0.190090597908551,2.18165690143323,17.3205080756888) {fixed-point};
%\node[fill=white, inner sep=0mm, rotate=286.07874135139, font=\bfseries, text=black!50!red, draw=black]
at (axis cs:0.881383322313993,0.0184366711899041,17.3205080756888) {LBFGS};
%\node[fill=white, inner sep=0mm, rotate=278.461835945856, font=\bfseries, text=black!50!green, draw=black]
at (axis cs:1.53369345545191,0.122208398635041,17.3205080756888) {CG};
%\node[fill=white, inner sep=0mm, rotate=280.891851926021, font=\bfseries, text=red!50!green, draw=black]
at (axis cs:2.20311301602805,180.065171636076,17.3205080756888) {TR};
\end{axis}
\end{tikzpicture}%
}
\caption{\small Comparison of the proposed fixed-point algorithms against manifold optimization techniques for EG distributions with dimension equal to 64 (left plot)  $a=1$ (right plot) $a=50$. FP1 and FP2 correspond to our proposed fixed-point algorithms for nonconcave and concave cases, respectively. K\&T correspond to the fixed point method proposed by~\citet{kent91}. LBFGS, CG and TR represent three manifold optimization methods. }
\label{fig.fp}
\end{figure}

We sampled 10,000 points from an EGD with a random scatter matrix, and initialized the iterations with a random positive definite matrix. The left plot in Fig.~\ref{fig.fp} shows the result for the case $a=1$ (nonconcave case) and right plot is for the case $a=50$ (concave case). For the fixed-point algorithm in nonconcave case, we use the scale parameter $\alpha_p$ as mentioned in Remark~\ref{rmk.1}.

\begin{figure}
\resizebox{.45\textwidth}{!}{\input{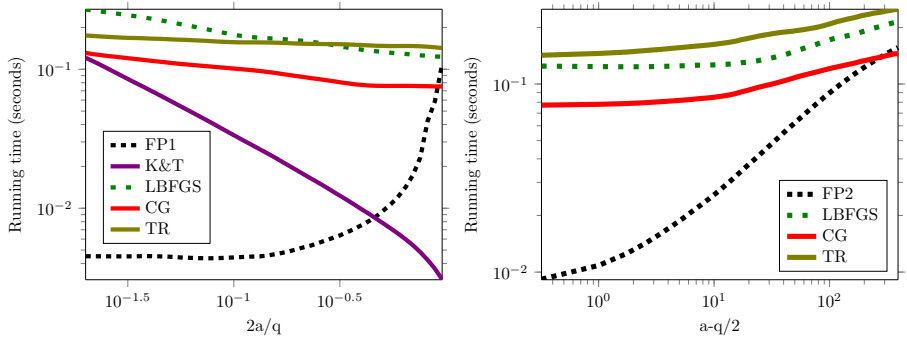}}
  \resizebox{.45\textwidth}{!}{% This file was created by matlab2tikz.
%
%The latest updates can be retrieved from
%  http://www.mathworks.com/matlabcentral/fileexchange/22022-matlab2tikz-matlab2tikz
%where you can also make suggestions and rate matlab2tikz.
%
\begin{tikzpicture}

\begin{axis}[%
scale=0.6,
width=4.822in,
height=3.661in,
at={(0.809in,0.656in)},
scale only axis,
separate axis lines,
every outer x axis line/.append style={black},
every x tick label/.append style={font=\color{black}},
xmode=log,
xmin=0.319998999999997,
xmax=392.000001,
tick align=outside,
xlabel={a-q/2},
every outer y axis line/.append style={black},
every y tick label/.append style={font=\color{black}},
ymode=log,
ymin=0.00910890073599999,
ymax=0.249837726618,
ylabel={Running time (seconds)},
axis background/.style={fill=white},
legend pos=south east,
legend style={legend cell align=left,align=left,draw=black}
]
\addplot [color=black,dashed,line width=3.0pt]
  table[row sep=crcr]{%
0.319999999999997	0.00916649263299999\\
1.0042711439082	0.0108613778946667\\
1.7448195712762	0.012545820066\\
2.54627375820122	0.014244106961\\
3.41364284576053	0.0159563210398\\
4.35234794747067	0.0176295726316\\
5.36825603160143	0.0193160934408\\
6.46771659011125	0.0210369414976\\
7.6576013233871	0.0227027585286\\
8.94534708882132	0.024412313834\\
10.3390023816566	0.0261994001318\\
11.8472776386075	0.028105191479\\
13.4795996786596	0.0300709340236\\
15.2461706213046	0.0321429857084\\
17.1580316504543	0.0343943977306\\
19.2271320225619	0.0366281454244\\
21.4664037502564	0.0388650854128\\
23.8898424282671	0.041182139005\\
26.5125947068059	0.0436911403444\\
29.3510529591216	0.0462475600196\\
32.4229577349046	0.0489033947746\\
35.7475086398825	0.0516596230078\\
39.3454843346116	0.0545455177576\\
43.2393724024641	0.0574972661514\\
47.4535098984945	0.0604335650206\\
52.0142354576252	0.0635461892714\\
56.9500539128375	0.0667689248512\\
62.2918144522409	0.0701656391450001\\
68.0729034285172	0.0735154074128\\
74.3294530258068	0.0770262608594\\
81.1005670882234	0.080596172146\\
88.4285655214359	0.084347963687\\
96.359248794847	0.0880573203764\\
104.942184197521	0.0918563330576\\
114.231015636984	0.0957297268922001\\
124.28379891715	0.0996617852132\\
135.163364590895	0.1035937730588\\
146.937710655121	0.1075703861132\\
159.680427542681	0.1116501079364\\
173.471158067399	0.1157170721124\\
188.396095196861	0.1198434890882\\
204.548520764104	0.1238693366356\\
222.029388485167	0.1279722206842\\
240.947954926424	0.1321349996288\\
261.422462365259	0.1362249744684\\
283.580877812043	0.1402945482978\\
307.561692812308	0.1444078114762\\
333.514789027967	0.1484125319816\\
361.602375007503	0.152381244367333\\
392	0.156357058436\\
};
\addlegendentry{FP2};

\addplot [color=black!50!green,loosely dashed,line width=3.0pt]
  table[row sep=crcr]{%
0.319999999999997	0.124572818864\\
1.0042711439082	0.123798823884667\\
1.7448195712762	0.1235401690554\\
2.54627375820122	0.1235656364276\\
3.41364284576053	0.1240718907446\\
4.35234794747067	0.124525643759\\
5.36825603160143	0.1250860298032\\
6.46771659011125	0.1255959746414\\
7.6576013233871	0.1261023873186\\
8.94534708882132	0.1260885348982\\
10.3390023816566	0.1265379480764\\
11.8472776386075	0.1271201220526\\
13.4795996786596	0.1276009601762\\
15.2461706213046	0.1281007325178\\
17.1580316504543	0.1298110025132\\
19.2271320225619	0.1305348698238\\
21.4664037502564	0.1315443351636\\
23.8898424282671	0.1330161422168\\
26.5125947068059	0.1348745755314\\
29.3510529591216	0.136222566154\\
32.4229577349046	0.1382056037816\\
35.7475086398825	0.1403287208584\\
39.3454843346116	0.1425745132002\\
43.2393724024641	0.1447000392958\\
47.4535098984945	0.1475447907892\\
52.0142354576252	0.1502735306738\\
56.9500539128375	0.1532783140232\\
62.2918144522409	0.156101765469\\
68.0729034285172	0.1590442315358\\
74.3294530258068	0.1615181673824\\
81.1005670882234	0.1644659192346\\
88.4285655214359	0.1672796438016\\
96.359248794847	0.1704205091108\\
104.942184197521	0.1733520951926\\
114.231015636984	0.1755123756438\\
124.28379891715	0.1775785872788\\
135.163364590895	0.1794069663264\\
146.937710655121	0.181368464398\\
159.680427542681	0.1829005351314\\
173.471158067399	0.1859256331586\\
188.396095196861	0.189050664236\\
204.548520764104	0.1919336256476\\
222.029388485167	0.1946112127632\\
240.947954926424	0.1983361826202\\
261.422462365259	0.2007527005448\\
283.580877812043	0.2036234936002\\
307.561692812308	0.2062514600984\\
333.514789027967	0.2094325284574\\
361.602375007503	0.212070587664\\
392	0.215586465126\\
};
\addlegendentry{LBFGS};

\addplot [color=red,solid,line width=3.0pt]
  table[row sep=crcr]{%
0.319999999999997	0.077245059907\\
1.0042711439082	0.078022703802\\
1.7448195712762	0.0787681458396\\
2.54627375820122	0.079591891967\\
3.41364284576053	0.0805407077418\\
4.35234794747067	0.081356613797\\
5.36825603160143	0.0821184335286001\\
6.46771659011125	0.0829495933028\\
7.6576013233871	0.0837840885606\\
8.94534708882132	0.0844049728722\\
10.3390023816566	0.0852828869248\\
11.8472776386075	0.0863360782952\\
13.4795996786596	0.0874400256068\\
15.2461706213046	0.0889047676286\\
17.1580316504543	0.0907705586376\\
19.2271320225619	0.0923757916778\\
21.4664037502564	0.0941059755578\\
23.8898424282671	0.0959128134006\\
26.5125947068059	0.0974183572374\\
29.3510529591216	0.0988299004088\\
32.4229577349046	0.100300236235\\
35.7475086398825	0.1019330297346\\
39.3454843346116	0.1035928722452\\
43.2393724024641	0.1053180187128\\
47.4535098984945	0.1071104020152\\
52.0142354576252	0.1089778194068\\
56.9500539128375	0.1106597793264\\
62.2918144522409	0.112192316887\\
68.0729034285172	0.113775151962\\
74.3294530258068	0.115269799566\\
81.1005670882234	0.1168520372424\\
88.4285655214359	0.118528174832\\
96.359248794847	0.1200727718274\\
104.942184197521	0.1216975889422\\
114.231015636984	0.1230636136972\\
124.28379891715	0.1244742261322\\
135.163364590895	0.1258673334128\\
146.937710655121	0.127344745751\\
159.680427542681	0.1286196385564\\
173.471158067399	0.1303472493942\\
188.396095196861	0.1321080393648\\
204.548520764104	0.1334962348902\\
222.029388485167	0.1349837876528\\
240.947954926424	0.136689543636\\
261.422462365259	0.1379317523408\\
283.580877812043	0.1395906091622\\
307.561692812308	0.1414066846954\\
333.514789027967	0.1428653886226\\
361.602375007503	0.144678580563667\\
392	0.144895325078\\
};
\addlegendentry{CG};

\addplot [color=red!50!green,solid,line width=3.0pt]
  table[row sep=crcr]{%
0.319999999999997	0.142276001288\\
1.0042711439082	0.145446581132333\\
1.7448195712762	0.1479110160894\\
2.54627375820122	0.1507238454216\\
3.41364284576053	0.152810330339\\
4.35234794747067	0.1549203054244\\
5.36825603160143	0.1566771094372\\
6.46771659011125	0.158449551395\\
7.6576013233871	0.1598988791702\\
8.94534708882132	0.1612712630822\\
10.3390023816566	0.162964535753\\
11.8472776386075	0.1648015387166\\
13.4795996786596	0.1666426942442\\
15.2461706213046	0.1688229059098\\
17.1580316504543	0.1720350834318\\
19.2271320225619	0.1743541095612\\
21.4664037502564	0.1768808781706\\
23.8898424282671	0.1794240429398\\
26.5125947068059	0.1817360112392\\
29.3510529591216	0.183481806901\\
32.4229577349046	0.1854982704428\\
35.7475086398825	0.187137433008\\
39.3454843346116	0.1886257746412\\
43.2393724024641	0.1897357009564\\
47.4535098984945	0.190849161348\\
52.0142354576252	0.1925446867936\\
56.9500539128375	0.1939218567598\\
62.2918144522409	0.1955304702418\\
68.0729034285172	0.197550851958\\
74.3294530258068	0.1992620089974\\
81.1005670882234	0.2014897470122\\
88.4285655214359	0.2042322659026\\
96.359248794847	0.207356616631\\
104.942184197521	0.2103986693618\\
114.231015636984	0.2141504947954\\
124.28379891715	0.2168075370924\\
135.163364590895	0.2196891156436\\
146.937710655121	0.2224350055166\\
159.680427542681	0.2246932780164\\
173.471158067399	0.2274876808712\\
188.396095196861	0.2308548177762\\
204.548520764104	0.233129585545\\
222.029388485167	0.2352700313122\\
240.947954926424	0.2380903450758\\
261.422462365259	0.2396643549286\\
283.580877812043	0.2415259404242\\
307.561692812308	0.244168125234\\
333.514789027967	0.2461714438918\\
361.602375007503	0.248444936946334\\
392	0.248852498823\\
};
\addlegendentry{TR};

\end{axis}
\end{tikzpicture}%
}
\caption{\small Effect of shape parameter on the convergence speed (left plot) nonconcave case (right case) concave case.}
\label{fig.faRnd}
\end{figure}

\begin{figure}
\resizebox{.45\textwidth}{!}{\input{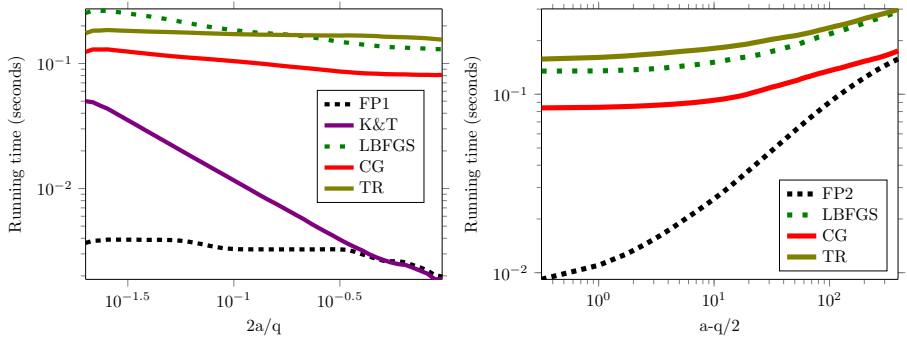}}
  \resizebox{.45\textwidth}{!}{% This file was created by matlab2tikz.
%
%The latest updates can be retrieved from
%  http://www.mathworks.com/matlabcentral/fileexchange/22022-matlab2tikz-matlab2tikz
%where you can also make suggestions and rate matlab2tikz.
%
\begin{tikzpicture}

\begin{axis}[%
scale=0.6,
width=4.822in,
height=3.661in,
at={(0.809in,0.656in)},
scale only axis,
separate axis lines,
every outer x axis line/.append style={black},
every x tick label/.append style={font=\color{black}},
xmode=log,
xmin=0.319998999999997,
xmax=392.000001,
tick align=outside,
xlabel={a-q/2},
every outer y axis line/.append style={black},
every y tick label/.append style={font=\color{black}},
ymode=log,
ymin=0.00916999119799999,
ymax=0.30100656432,
ylabel={Running time (seconds)},
axis background/.style={fill=white},
legend pos=south east,
legend style={legend cell align=left,align=left,draw=black}
]
\addplot [color=black,dashed,line width=3.0pt]
  table[row sep=crcr]{%
0.319999999999997	0.00918634624\\
1.0042711439082	0.011009773776\\
1.7448195712762	0.0127286042092\\
2.54627375820122	0.0144559828112\\
3.41364284576053	0.0161448529726\\
4.35234794747067	0.0178329190212\\
5.36825603160143	0.0195826037512\\
6.46771659011125	0.0212419390488\\
7.6576013233871	0.0229149271212\\
8.94534708882132	0.0246009048404\\
10.3390023816566	0.026367268451\\
11.8472776386075	0.0281473113656\\
13.4795996786596	0.0301543852378\\
15.2461706213046	0.0322037696652\\
17.1580316504543	0.034343837383\\
19.2271320225619	0.0366121070386\\
21.4664037502564	0.038891309746\\
23.8898424282671	0.0413106054294\\
26.5125947068059	0.0437576605102\\
29.3510529591216	0.0463375749148\\
32.4229577349046	0.048974589708\\
35.7475086398825	0.0517621706398\\
39.3454843346116	0.0545800317846\\
43.2393724024641	0.0575655126316\\
47.4535098984945	0.0606194338104\\
52.0142354576252	0.0638091602466\\
56.9500539128375	0.0670675113144\\
62.2918144522409	0.0704458579612\\
68.0729034285172	0.0739032100426\\
74.3294530258068	0.0774158181054\\
81.1005670882234	0.0810051467252\\
88.4285655214359	0.0847087109462\\
96.359248794847	0.0884722204100001\\
104.942184197521	0.0922388995120001\\
114.231015636984	0.096251698908\\
124.28379891715	0.1001959232698\\
135.163364590895	0.1040611958564\\
146.937710655121	0.1081531569664\\
159.680427542681	0.1122934542972\\
173.471158067399	0.116334668714\\
188.396095196861	0.1205851745744\\
204.548520764104	0.1249021254726\\
222.029388485167	0.1290057290644\\
240.947954926424	0.1332509979326\\
261.422462365259	0.1375313865184\\
283.580877812043	0.141498349638\\
307.561692812308	0.1456358271248\\
333.514789027967	0.1497120096768\\
361.602375007503	0.153584299100333\\
392	0.157712255752\\
};
\addlegendentry{FP2};

\addplot [color=black!50!green,loosely dashed,line width=3.0pt]
  table[row sep=crcr]{%
0.319999999999997	0.134848658794\\
1.0042711439082	0.135148942914334\\
1.7448195712762	0.1366701363006\\
2.54627375820122	0.138123761073\\
3.41364284576053	0.1397581262978\\
4.35234794747067	0.1418501984986\\
5.36825603160143	0.1439403231436\\
6.46771659011125	0.1456642393142\\
7.6576013233871	0.1475548117808\\
8.94534708882132	0.1495680354852\\
10.3390023816566	0.1517772395426\\
11.8472776386075	0.1538238384434\\
13.4795996786596	0.1562327070154\\
15.2461706213046	0.158779045838\\
17.1580316504543	0.1603686332934\\
19.2271320225619	0.1618189051422\\
21.4664037502564	0.1644363863136\\
23.8898424282671	0.1662364672862\\
26.5125947068059	0.1682368358686\\
29.3510529591216	0.17138719634\\
32.4229577349046	0.1746925544988\\
35.7475086398825	0.1768478799024\\
39.3454843346116	0.180220243033\\
43.2393724024641	0.1835062031978\\
47.4535098984945	0.1863102965596\\
52.0142354576252	0.1888316342624\\
56.9500539128375	0.1935261541644\\
62.2918144522409	0.1976558391114\\
68.0729034285172	0.2010910049978\\
74.3294530258068	0.205161336925\\
81.1005670882234	0.2098208333308\\
88.4285655214359	0.2125350176188\\
96.359248794847	0.2159566658088\\
104.942184197521	0.2197798345438\\
114.231015636984	0.223767823212\\
124.28379891715	0.2267589924354\\
135.163364590895	0.2306018987986\\
146.937710655121	0.2348290949438\\
159.680427542681	0.2389402393954\\
173.471158067399	0.2429566375578\\
188.396095196861	0.247406905726\\
204.548520764104	0.2523186876934\\
222.029388485167	0.2561648131944\\
240.947954926424	0.2615948041438\\
261.422462365259	0.2651885726836\\
283.580877812043	0.2697348479186\\
307.561692812308	0.2745918107488\\
333.514789027967	0.2801521880442\\
361.602375007503	0.285336056031667\\
392	0.293369504012\\
};
\addlegendentry{LBFGS};

\addplot [color=red,solid,line width=3.0pt]
  table[row sep=crcr]{%
0.319999999999997	0.083854857632\\
1.0042711439082	0.08450412187\\
1.7448195712762	0.0853639838092\\
2.54627375820122	0.0862098540566\\
3.41364284576053	0.08706806172\\
4.35234794747067	0.0879194001244\\
5.36825603160143	0.0888003382818\\
6.46771659011125	0.0896556557194\\
7.6576013233871	0.0906898662584\\
8.94534708882132	0.0917148160524\\
10.3390023816566	0.0928522667888001\\
11.8472776386075	0.0938521049928\\
13.4795996786596	0.0952526870932\\
15.2461706213046	0.096576991313\\
17.1580316504543	0.0980345145852\\
19.2271320225619	0.0997910005143999\\
21.4664037502564	0.1020004916592\\
23.8898424282671	0.1038577805202\\
26.5125947068059	0.1059153689618\\
29.3510529591216	0.1080669946668\\
32.4229577349046	0.1100311497028\\
35.7475086398825	0.1116525728848\\
39.3454843346116	0.1136868074498\\
43.2393724024641	0.1157501350372\\
47.4535098984945	0.1176552088532\\
52.0142354576252	0.1195303360468\\
56.9500539128375	0.1221147650016\\
62.2918144522409	0.1243506631632\\
68.0729034285172	0.1263200298904\\
74.3294530258068	0.128557071009\\
81.1005670882234	0.130758445191\\
88.4285655214359	0.1325408774468\\
96.359248794847	0.1345569119568\\
104.942184197521	0.136597649648\\
114.231015636984	0.1386602494938\\
124.28379891715	0.1403488083708\\
135.163364590895	0.1424686525126\\
146.937710655121	0.144886530121\\
159.680427542681	0.1469156051192\\
173.471158067399	0.149209702912\\
188.396095196861	0.1514477838716\\
204.548520764104	0.1536864997072\\
222.029388485167	0.1554255153358\\
240.947954926424	0.157990518751\\
261.422462365259	0.1597501100016\\
283.580877812043	0.1625067437788\\
307.561692812308	0.165103110592\\
333.514789027967	0.1681705932396\\
361.602375007503	0.171385076071333\\
392	0.175142415798\\
};
\addlegendentry{CG};

\addplot [color=red!50!green,solid,line width=3.0pt]
  table[row sep=crcr]{%
0.319999999999997	0.157477226117\\
1.0042711439082	0.160869853439667\\
1.7448195712762	0.1640264082878\\
2.54627375820122	0.1671427248318\\
3.41364284576053	0.169464817822\\
4.35234794747067	0.1720571614122\\
5.36825603160143	0.1742135538354\\
6.46771659011125	0.1759074721986\\
7.6576013233871	0.177629719167\\
8.94534708882132	0.1795295972906\\
10.3390023816566	0.181481667889\\
11.8472776386075	0.183125386537\\
13.4795996786596	0.1851206127496\\
15.2461706213046	0.1871455315152\\
17.1580316504543	0.1891419237554\\
19.2271320225619	0.1910603316362\\
21.4664037502564	0.1938378916088\\
23.8898424282671	0.1962864595244\\
26.5125947068059	0.1987028468578\\
29.3510529591216	0.2013110313918\\
32.4229577349046	0.203757570076\\
35.7475086398825	0.2056947257992\\
39.3454843346116	0.207994266855\\
43.2393724024641	0.2099725150134\\
47.4535098984945	0.2117290220914\\
52.0142354576252	0.2136615230672\\
56.9500539128375	0.2164631814232\\
62.2918144522409	0.2190308727934\\
68.0729034285172	0.22175599315\\
74.3294530258068	0.2248353500538\\
81.1005670882234	0.2283425233378\\
88.4285655214359	0.231441768823\\
96.359248794847	0.2345399531148\\
104.942184197521	0.2377607177796\\
114.231015636984	0.2412073996498\\
124.28379891715	0.2439361163442\\
135.163364590895	0.2470971684656\\
146.937710655121	0.250490752289\\
159.680427542681	0.2543075982498\\
173.471158067399	0.2582696733514\\
188.396095196861	0.262518149922\\
204.548520764104	0.2668847842882\\
222.029388485167	0.270453436141\\
240.947954926424	0.2744074868108\\
261.422462365259	0.2774541695476\\
283.580877812043	0.281507615514\\
307.561692812308	0.2853097344824\\
333.514789027967	0.289617946056\\
361.602375007503	0.294324343131667\\
392	0.297702223729\\
};
\addlegendentry{TR};

\end{axis}
\end{tikzpicture}%
}
\caption{\small Effect of shape parameter on the convergence speed where sample covariance is used for initialization.}
\label{fig.faSam}
\end{figure}

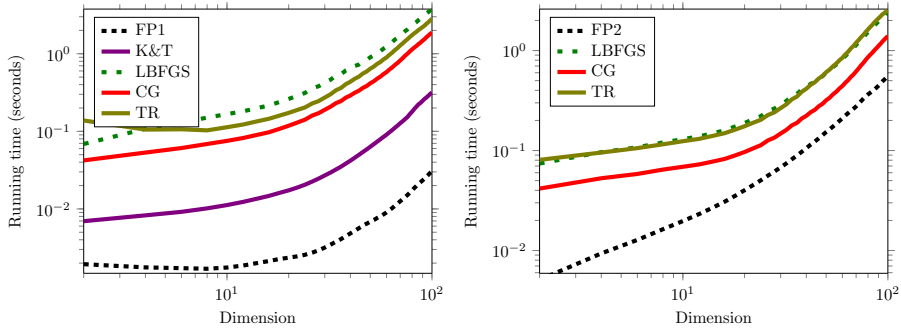
\begin{figure}
  \resizebox{.45\textwidth}{!}{% This file was created by matlab2tikz.
%
%The latest updates can be retrieved from
%  http://www.mathworks.com/matlabcentral/fileexchange/22022-matlab2tikz-matlab2tikz
%where you can also make suggestions and rate matlab2tikz.
%
\begin{tikzpicture}

\begin{axis}[%
scale=0.6,
width=4.822in,
height=3.661in,
at={(0.809in,0.656in)},
scale only axis,
separate axis lines,
every outer x axis line/.append style={black},
every x tick label/.append style={font=\color{black}},
xmode=log,
xmin=1.999999,
xmax=100.000001,
tick align=outside,
xlabel={Dimension},
every outer y axis line/.append style={black},
every y tick label/.append style={font=\color{black}},
ymode=log,
ymin=0.001478757566,
ymax=3.76515293916,
ylabel={Running time (seconds)},
axis background/.style={fill=white},
legend pos=north west,
legend style={legend cell align=left,align=left,draw=black}
]
\addplot [color=black,dashed,line width=2.5pt]
  table[row sep=crcr]{%
2	0.001938900414\\
4	0.00176258943\\
6	0.0017220301866\\
8	0.001692189145\\
10	0.0017544245852\\
12	0.0018633129588\\
14	0.0020018566402\\
16	0.0021260073848\\
18	0.0022593919394\\
20	0.0023498788108\\
22	0.0024454499116\\
24	0.002562612651\\
26	0.002726968753\\
28	0.002924013786\\
30	0.0031744944402\\
32	0.003463415128\\
34	0.0037687360894\\
36	0.0040858566382\\
38	0.0044551410586\\
40	0.0048356695732\\
42	0.0052221722734\\
44	0.0056395375332\\
46	0.006078821902\\
48	0.0064498031236\\
50	0.0068310611294\\
52	0.007184815764\\
54	0.0076357695938\\
56	0.0080097824716\\
58	0.0084940934214\\
60	0.00900985198020001\\
62	0.0096255907734\\
64	0.0102436541738\\
66	0.0109837743426\\
68	0.0117323934698\\
70	0.0125600696986\\
72	0.0133615423666\\
74	0.014228865389\\
76	0.015131943839\\
78	0.0162395155516\\
80	0.0173491582954\\
82	0.0186215305826\\
84	0.0197543013436\\
86	0.0210471815774\\
88	0.022144911638\\
90	0.0233213827546\\
92	0.0245547711396\\
94	0.0260745459688\\
96	0.0274944109302\\
98	0.0291542386773333\\
100	0.030586028891\\
};
\addlegendentry{FP1};

\addplot [color=violet,solid,line width=2.5pt]
  table[row sep=crcr]{%
2	0.006944331731\\
4	0.00825762653566667\\
6	0.0091671809616\\
8	0.0101658931372\\
10	0.0112038239716\\
12	0.0123178739986\\
14	0.0135367179994\\
16	0.0146985392778\\
18	0.0160610935536\\
20	0.0173473063238\\
22	0.0188273507836\\
24	0.0205030580934\\
26	0.022476337156\\
28	0.0246001879114\\
30	0.026996567668\\
32	0.0294368162464\\
34	0.0322200859466\\
36	0.0350769735376\\
38	0.0384228544282\\
40	0.0418412671566\\
42	0.045692246084\\
44	0.0497561596422\\
46	0.0540513129012\\
48	0.058439914244\\
50	0.0632681724598\\
52	0.0682853138998\\
54	0.073548563307\\
56	0.0790343347174\\
58	0.084735235131\\
60	0.0905671816048\\
62	0.097198909933\\
64	0.1039315439378\\
66	0.1114165342218\\
68	0.1188317772068\\
70	0.1273014995924\\
72	0.135055711727\\
74	0.1442119155338\\
76	0.1532117081126\\
78	0.167064791736\\
80	0.1792427753144\\
82	0.1943263930322\\
84	0.208448895596\\
86	0.2229173574196\\
88	0.2333269405216\\
90	0.2473414755258\\
92	0.2617097965068\\
94	0.2734497006038\\
96	0.2879453778618\\
98	0.303005754843333\\
100	0.317371056532\\
};
\addlegendentry{K\&T};

\addplot [color=black!50!green,loosely dashed,line width=2.5pt]
  table[row sep=crcr]{%
2	0.068690026209\\
4	0.107578319760667\\
6	0.1281859442022\\
8	0.1505526437368\\
10	0.166617211124\\
12	0.181367964534\\
14	0.198995079986\\
16	0.21537389428\\
18	0.2382801309482\\
20	0.2612532456358\\
22	0.2876379778672\\
24	0.3122842285928\\
26	0.3574593985336\\
28	0.381607554711\\
30	0.4235267400686\\
32	0.4626278984432\\
34	0.5219280716376\\
36	0.5471059091516\\
38	0.6076476754672\\
40	0.641958151082\\
42	0.6954130716642\\
44	0.72554786966\\
46	0.7885105043876\\
48	0.8301016023902\\
50	0.9001434448862\\
52	0.9499502745002\\
54	1.0230489004816\\
56	1.0809218593884\\
58	1.1629777156524\\
60	1.2295953378578\\
62	1.3305169658106\\
64	1.41539038354\\
66	1.5254743726246\\
68	1.6138945633908\\
70	1.7311016149748\\
72	1.8293641426042\\
74	1.9591870297616\\
76	2.0620354466686\\
78	2.202815921582\\
80	2.3234162707106\\
82	2.4578276902456\\
84	2.567721365356\\
86	2.713419623653\\
88	2.8401139157284\\
90	2.984044629643\\
92	3.1188975229436\\
94	3.293025307646\\
96	3.44493436219\\
98	3.60751275962633\\
100	3.760071804841\\
};
\addlegendentry{LBFGS};

\addplot [color=red,solid,line width=2.5pt]
  table[row sep=crcr]{%
2	0.042178502839\\
4	0.0532369930943334\\
6	0.0608824212386\\
8	0.0686785660068\\
10	0.075337685046\\
12	0.0820595575522\\
14	0.0898898583716\\
16	0.0971874233208\\
18	0.1074213203182\\
20	0.1181465205678\\
22	0.12961758901\\
24	0.1412583573588\\
26	0.1594704776304\\
28	0.1715402783068\\
30	0.1896149211384\\
32	0.2077377133282\\
34	0.2307827426742\\
36	0.245595063183\\
38	0.2720071223532\\
40	0.2897958569272\\
42	0.314551161553\\
44	0.33398986387\\
46	0.3626276486428\\
48	0.3840480406066\\
50	0.4160528775678\\
52	0.441695654539\\
54	0.4768322891718\\
56	0.5066579810884\\
58	0.5467542548272\\
60	0.5801543423826\\
62	0.6261518023648\\
64	0.6665379355296\\
66	0.7189695800714\\
68	0.7629950947808\\
70	0.8201651471548\\
72	0.8779572627002\\
74	0.9453254723882\\
76	1.0028945059982\\
78	1.0738009093222\\
80	1.1381231947106\\
82	1.2028310499588\\
84	1.2606018878424\\
86	1.333552014021\\
88	1.3987166844538\\
90	1.4728417341386\\
92	1.5452190948756\\
94	1.6358042623298\\
96	1.713840568658\\
98	1.80334015204533\\
100	1.875932084324\\
};
\addlegendentry{CG};

\addplot [color=red!50!green,solid,line width=2.5pt]
  table[row sep=crcr]{%
2	0.137913970251\\
4	0.105312063831667\\
6	0.1062283224294\\
8	0.1029415652966\\
10	0.112956882068\\
12	0.122450648045\\
14	0.1343641853354\\
16	0.1451014574502\\
18	0.1591685026956\\
20	0.1730008252436\\
22	0.1884355389596\\
24	0.2030266891326\\
26	0.2283500662818\\
28	0.2452810834768\\
30	0.2726411391776\\
32	0.3003438170986\\
34	0.3382395438538\\
36	0.3626325497134\\
38	0.4057581169268\\
40	0.435876507288199\\
42	0.4762132322196\\
44	0.5066842526472\\
46	0.5552550381028\\
48	0.5899652499034\\
50	0.6420571155472\\
52	0.6830659925278\\
54	0.7391081248632\\
56	0.78278766332\\
58	0.84507515664\\
60	0.8953976948552\\
62	0.9690132297244\\
64	1.0339656847788\\
66	1.1178193116128\\
68	1.187826514855\\
70	1.2801954522018\\
72	1.36106990291\\
74	1.4579910259582\\
76	1.5410186539424\\
78	1.644747828936\\
80	1.7382383861562\\
82	1.8375746985434\\
84	1.9217684633228\\
86	2.02688659481\\
88	2.1118428158982\\
90	2.2059257835752\\
92	2.3025322652488\\
94	2.434377663902\\
96	2.5501744662402\\
98	2.692980692143\\
100	2.816753800088\\
};
\addlegendentry{TR};

\end{axis}
\end{tikzpicture}%
}
  \resizebox{.45\textwidth}{!}{% This file was created by matlab2tikz.
%
%The latest updates can be retrieved from
%  http://www.mathworks.com/matlabcentral/fileexchange/22022-matlab2tikz-matlab2tikz
%where you can also make suggestions and rate matlab2tikz.
%
\begin{tikzpicture}

\begin{axis}[%
scale=0.6,
width=4.822in,
height=3.661in,
at={(0.809in,0.656in)},
scale only axis,
separate axis lines,
every outer x axis line/.append style={black},
every x tick label/.append style={font=\color{black}},
xmode=log,
xmin=1.999999,
xmax=100.000001,
tick align=outside,
xlabel={Dimension},
every outer y axis line/.append style={black},
every y tick label/.append style={font=\color{black}},
ymode=log,
ymin=0.005915511654,
ymax=2.598950192692,
ylabel={Running time (seconds)},
axis background/.style={fill=white},
legend pos=north west,
legend style={legend cell align=left,align=left,draw=black}
]
\addplot [color=black,dashed,line width=2.5pt]
  table[row sep=crcr]{%
2	0.00508479848499999\\
4	0.00935099767133333\\
6	0.0127625887066\\
8	0.0163437159712\\
10	0.0196997857968\\
12	0.0231420783442\\
14	0.0268699930106\\
16	0.0307905658956\\
18	0.035266888853\\
20	0.0401085619696\\
22	0.0451042090548\\
24	0.0503716151068\\
26	0.0561442560082\\
28	0.0617493853388\\
30	0.0679506704002\\
32	0.0744410236328\\
34	0.0817329023658\\
36	0.0891326629318\\
38	0.0970782832656\\
40	0.105412652601\\
42	0.1145799420752\\
44	0.1229521261916\\
46	0.1323269894348\\
48	0.1425270002196\\
50	0.1530444568334\\
52	0.1627148067418\\
54	0.173870286638\\
56	0.1852429981872\\
58	0.1975865125404\\
60	0.2102254878558\\
62	0.2239387260452\\
64	0.236468634716\\
66	0.2492913733246\\
68	0.2620302764392\\
70	0.2760199282068\\
72	0.2927534414434\\
74	0.3116365504982\\
76	0.3305215171282\\
78	0.3482980729874\\
80	0.3641896126874\\
82	0.3782038321228\\
84	0.3922400566152\\
86	0.4078840450942\\
88	0.4255348555944\\
90	0.4454777997704\\
92	0.467064826166\\
94	0.4902217287174\\
96	0.5109331103532\\
98	0.533234644458667\\
100	0.549222230279\\
};
\addlegendentry{FP2};

\addplot [color=black!50!green,loosely dashed,line width=2.5pt]
  table[row sep=crcr]{%
2	0.074171919338\\
4	0.096231118553\\
6	0.1078508862264\\
8	0.1205763644652\\
10	0.1297977694496\\
12	0.1386154735098\\
14	0.148813141341\\
16	0.1594799120604\\
18	0.173921623707\\
20	0.1884771652474\\
22	0.2055885707138\\
24	0.220542241067\\
26	0.2487402099694\\
28	0.2629435874952\\
30	0.289472487163\\
32	0.3132445161494\\
34	0.3520648807932\\
36	0.3677081055212\\
38	0.4077445997738\\
40	0.429062557553\\
42	0.4646263421732\\
44	0.4838341188222\\
46	0.5256107159588\\
48	0.5535743723714\\
50	0.602138353147\\
52	0.6348314843986\\
54	0.684691302972\\
56	0.722202057964\\
58	0.7754162006224\\
60	0.8165960106814\\
62	0.8807586275022\\
64	0.933868103711\\
66	1.0075422663292\\
68	1.0628526776092\\
70	1.1385743315572\\
72	1.2060032320636\\
74	1.2911373131432\\
76	1.3562745926034\\
78	1.4475847790584\\
80	1.5289357105064\\
82	1.6172086458992\\
84	1.6893024072084\\
86	1.7841007259516\\
88	1.8660536162198\\
90	1.9580835044378\\
92	2.0408782753058\\
94	2.143011377574\\
96	2.2314427315124\\
98	2.32448706533933\\
100	2.409293363283\\
};
\addlegendentry{LBFGS};

\addplot [color=red,solid,line width=2.5pt]
  table[row sep=crcr]{%
2	0.041632716578\\
4	0.0528033149753334\\
6	0.058300434096\\
8	0.0644645893478\\
10	0.0687684900358\\
12	0.0726134899686\\
14	0.0775868139026\\
16	0.082448647945\\
18	0.0893592925458\\
20	0.0966331000694001\\
22	0.1049535267752\\
24	0.1130849291868\\
26	0.1263314297264\\
28	0.134803819912\\
30	0.1480161228944\\
32	0.1609475150886\\
34	0.1784900305792\\
36	0.1890751296702\\
38	0.2089572638302\\
40	0.222283216793\\
42	0.2415154516914\\
44	0.2550278581216\\
46	0.2770040820064\\
48	0.2933203160808\\
50	0.3181339272986\\
52	0.3373284295892\\
54	0.3649087308978\\
56	0.3872424639584\\
58	0.4169026581608\\
60	0.442024295806\\
62	0.4761739199906\\
64	0.5058353460022\\
66	0.544589538957\\
68	0.577146521328\\
70	0.618739043905\\
72	0.6612029856052\\
74	0.7107743427368\\
76	0.7543356030374\\
78	0.808415000664\\
80	0.8581925854626\\
82	0.9069733265084\\
84	0.952106710595\\
86	1.0072489684496\\
88	1.0559159352264\\
90	1.1114855333124\\
92	1.1655864038752\\
94	1.2291372362494\\
96	1.2832840806636\\
98	1.343653607436\\
100	1.39223913377\\
};
\addlegendentry{CG};

\addplot [color=red!50!green,solid,line width=2.5pt]
  table[row sep=crcr]{%
2	0.0806609908690001\\
4	0.0954853784416667\\
6	0.105722405415\\
8	0.1154069141774\\
10	0.1239253397322\\
12	0.1308228673078\\
14	0.1401647569402\\
16	0.149166735158\\
18	0.1617289528628\\
20	0.1749267484066\\
22	0.1895350156768\\
24	0.2028845575066\\
26	0.2264748733394\\
28	0.2413587295278\\
30	0.2658990867348\\
32	0.290051210398\\
34	0.3250503263154\\
36	0.346262362018\\
38	0.3869132979334\\
40	0.41486993063\\
42	0.4535968222846\\
44	0.4814631651798\\
46	0.5257210488724\\
48	0.5578977123508\\
50	0.60730086572\\
52	0.6447755028958\\
54	0.6969463077016\\
56	0.740539925521\\
58	0.7986870027876\\
60	0.8451218782298\\
62	0.9134484747234\\
64	0.9727701297546\\
66	1.0497666825388\\
68	1.1136688454556\\
70	1.1966724874308\\
72	1.2722095362454\\
74	1.360652808184\\
76	1.4369441390932\\
78	1.5357758032508\\
80	1.625695345744\\
82	1.7198485461532\\
84	1.8078782555978\\
86	1.917917714874\\
88	2.005611523966\\
90	2.1010554104258\\
92	2.1903131457992\\
94	2.2988963419014\\
96	2.370458351714\\
98	2.464217274621\\
100	2.508720340852\\
};
\addlegendentry{TR};

\end{axis}
\end{tikzpicture}%
}
\caption{\small Effect of dimension on the convergence speed.}
\label{fig.fdSam}
\end{figure}

\begin{figure}
  \resizebox{.45\textwidth}{!}{% This file was created by matlab2tikz.
%
%The latest updates can be retrieved from
%  http://www.mathworks.com/matlabcentral/fileexchange/22022-matlab2tikz-matlab2tikz
%where you can also make suggestions and rate matlab2tikz.
%
\begin{tikzpicture}

\begin{axis}[%
scale=0.6,
width=4.822in,
height=3.661in,
at={(0.809in,0.656in)},
scale only axis,
separate axis lines,
every outer x axis line/.append style={black},
every x tick label/.append style={font=\color{black}},
xmode=log,
xmin=19.999999,
xmax=16000.000001,
tick align=outside,
xlabel={Number of samples},
every outer y axis line/.append style={black},
every y tick label/.append style={font=\color{black}},
ymode=log,
ymin=0.002049828352,
ymax=10.204401808096,
ylabel={Running time (seconds)},
axis background/.style={fill=white},
legend pos=north west,
legend style={legend cell align=left,align=left,draw=black}
]
\addplot [color=black,dashed,line width=2.5pt]
  table[row sep=crcr]{%
20	0.007047458492\\
23	0.00528019056833333\\
26	0.004409806166\\
30	0.0036092432714\\
35	0.0032031724454\\
40	0.0029388698984\\
45	0.0027651848566\\
52	0.0026617737388\\
60	0.0025088378294\\
68	0.0023875100758\\
78	0.0022993635088\\
90	0.0022266319176\\
103	0.0021681912812\\
118	0.0021305806782\\
135	0.0021053787104\\
155	0.0020891701228\\
177	0.0021001097442\\
203	0.0021305720082\\
233	0.0021785839026\\
267	0.002231104551\\
306	0.00227860779\\
351	0.0023414848868\\
402	0.0024232367166\\
461	0.0025293004706\\
528	0.0026780923882\\
606	0.0028845623386\\
694	0.0031225498444\\
796	0.0034121178574\\
912	0.003704761089\\
1045	0.003979218156\\
1198	0.0042405491932\\
1373	0.0045399014418\\
1574	0.004867588361\\
1804	0.005282527219\\
2067	0.0058329546322\\
2370	0.0065150529436\\
2716	0.0073191834016\\
3113	0.0082583937324\\
3568	0.009352363437\\
4089	0.0106184581438\\
4687	0.0120906073476\\
5372	0.0137979769628\\
6157	0.0157189994522\\
7057	0.0179198722958\\
8089	0.0202292110776\\
9271	0.022923230974\\
10626	0.025945921914\\
12179	0.029670631257\\
13960	0.0334602233863333\\
16000	0.038651391773\\
};
\addlegendentry{FP1};

\addplot [color=violet,solid,line width=2.5pt]
  table[row sep=crcr]{%
20	0.01671964217\\
23	0.015161211517\\
26	0.014429838078\\
30	0.0141144161066\\
35	0.014166606635\\
40	0.0143766166776\\
45	0.0146815816598\\
52	0.015032698662\\
60	0.0150017121288\\
68	0.0149365088718\\
78	0.0148416767292\\
90	0.0147652310606\\
103	0.0146845252506\\
118	0.0146383482048\\
135	0.0146858095752\\
155	0.0148171388248\\
177	0.0150304284068\\
203	0.0153097901622\\
233	0.015641182517\\
267	0.0160299043892\\
306	0.0164752965814\\
351	0.0169407886804\\
402	0.0175101769998\\
461	0.0181621957744\\
528	0.0188573354414\\
606	0.0195814834324\\
694	0.0203511462354\\
796	0.0212427755778\\
912	0.022099861628\\
1045	0.0229870103352\\
1198	0.023881826345\\
1373	0.0247367840446\\
1574	0.0253799964638\\
1804	0.0259479301874\\
2067	0.0262893176574\\
2370	0.0263388791846\\
2716	0.0260029682872\\
3113	0.0252983342076\\
3568	0.0240079647352\\
4089	0.0222132575522\\
4687	0.0203400478354\\
5372	0.0189672876636\\
6157	0.0183540507442\\
7057	0.0188689120372\\
8089	0.0203586695716\\
9271	0.02278390019\\
10626	0.0257723678864\\
12179	0.0294399470598\\
13960	0.033093477655\\
16000	0.038273387404\\
};
\addlegendentry{K\&T};

\addplot [color=black!50!green,loosely dashed,line width=2.5pt]
  table[row sep=crcr]{%
20	0.232527756484\\
23	0.223054546297\\
26	0.2200177949286\\
30	0.2172404831526\\
35	0.216528235626\\
40	0.2164784673574\\
45	0.216977013637\\
52	0.2163749700988\\
60	0.2154238760522\\
68	0.2149251027582\\
78	0.2143044496062\\
90	0.2137814657286\\
103	0.2133273487734\\
118	0.2132712417072\\
135	0.2127976819528\\
155	0.212797937354\\
177	0.212470116453\\
203	0.2122628622874\\
233	0.2118178823452\\
267	0.2117232245462\\
306	0.2118825615736\\
351	0.2121077597264\\
402	0.2131702440274\\
461	0.2143620629774\\
528	0.2158364628344\\
606	0.217107128911\\
694	0.219276021201\\
796	0.2214180793692\\
912	0.224251090035\\
1045	0.2274317963612\\
1198	0.2314306516942\\
1373	0.2359337808818\\
1574	0.2410453859674\\
1804	0.2468666645454\\
2067	0.2536429089136\\
2370	0.2613220676268\\
2716	0.2698943392288\\
3113	0.2801958965914\\
3568	0.2916542347154\\
4089	0.3047941614242\\
4687	0.3189257597458\\
5372	0.3357503647316\\
6157	0.3550861807666\\
7057	0.3771761897802\\
8089	0.402142059666\\
9271	0.4464501421988\\
10626	0.4806757665032\\
12179	0.537304484585\\
13960	0.599704883506333\\
16000	0.683971326142\\
};
\addlegendentry{LBFGS};

\addplot [color=red,solid,line width=2.5pt]
  table[row sep=crcr]{%
20	0.103879557222\\
23	0.0991816956699999\\
26	0.097771544281\\
30	0.0963648276164\\
35	0.0961481577918\\
40	0.0964521012964001\\
45	0.0970243309554001\\
52	0.0967819916770001\\
60	0.0967289696524001\\
68	0.0966533893164\\
78	0.0963411476366\\
90	0.096212510471\\
103	0.0962442067836\\
118	0.096169126488\\
135	0.0963055843858\\
155	0.096793494615\\
177	0.0973185711552\\
203	0.0977095739846\\
233	0.098263165519\\
267	0.0987460735174\\
306	0.0992532896168\\
351	0.1000930335532\\
402	0.1010292686736\\
461	0.1024311748686\\
528	0.1041181727502\\
606	0.1060997038404\\
694	0.1080863309384\\
796	0.11113389572\\
912	0.114513412422\\
1045	0.1184638207316\\
1198	0.1230182986066\\
1373	0.1283063624854\\
1574	0.1341105136494\\
1804	0.1406914289278\\
2067	0.1482452311782\\
2370	0.1569163918452\\
2716	0.1669576874068\\
3113	0.1789209027356\\
3568	0.1919679251012\\
4089	0.206920465754\\
4687	0.2238461877326\\
5372	0.2437295164588\\
6157	0.2662013762636\\
7057	0.292208316596\\
8089	0.3200676747658\\
9271	0.3663349819884\\
10626	0.403345864847\\
12179	0.4631317288664\\
13960	0.528334233388333\\
16000	0.618515285992998\\
};
\addlegendentry{CG};

\addplot [color=red!50!green,solid,line width=2.5pt]
  table[row sep=crcr]{%
20	0.165353879556\\
23	0.162878783805\\
26	0.1614460537418\\
30	0.1601182555332\\
35	0.158945340467\\
40	0.1577614975818\\
45	0.1566309055258\\
52	0.1546118605956\\
60	0.1527346707726\\
68	0.1507725369742\\
78	0.1489450576564\\
90	0.1476179888384\\
103	0.1463613857834\\
118	0.1455254943858\\
135	0.1446969532848\\
155	0.144562712619\\
177	0.144434061417\\
203	0.1444790779168\\
233	0.144613452232\\
267	0.1456027197136\\
306	0.146554448235\\
351	0.1481635469504\\
402	0.1502410560588\\
461	0.153033552146\\
528	0.1562404292114\\
606	0.1600437901424\\
694	0.1644725770268\\
796	0.1702381211934\\
912	0.1768929007638\\
1045	0.1845343270772\\
1198	0.1930993438314\\
1373	0.2031798262404\\
1574	0.2144088889248\\
1804	0.2272264914958\\
2067	0.2421195884074\\
2370	0.2594010447284\\
2716	0.278975697331\\
3113	0.3018688634512\\
3568	0.3272935147342\\
4089	0.3563981946622\\
4687	0.3901955974138\\
5372	0.4284575457546\\
6157	0.4725019397106\\
7057	0.5225555247404\\
8089	0.5797602271938\\
9271	0.677128226516201\\
10626	0.754201776157\\
12179	0.8779868336822\\
13960	1.01585582712667\\
16000	1.193941858955\\
};
\addlegendentry{TR};

\end{axis}
\end{tikzpicture}%
}
  \resizebox{.45\textwidth}{!}{% This file was created by matlab2tikz.
%
%The latest updates can be retrieved from
%  http://www.mathworks.com/matlabcentral/fileexchange/22022-matlab2tikz-matlab2tikz
%where you can also make suggestions and rate matlab2tikz.
%
\begin{tikzpicture}

\begin{axis}[%
scale=0.6,
width=4.822in,
height=3.661in,
at={(0.809in,0.656in)},
scale only axis,
separate axis lines,
every outer x axis line/.append style={black},
every x tick label/.append style={font=\color{black}},
xmode=log,
xmin=19.999999,
xmax=16000.000001,
tick align=outside,
xlabel={Number of samples},
every outer y axis line/.append style={black},
every y tick label/.append style={font=\color{black}},
ymode=log,
ymin=0.02176137464,
ymax=1.09512241757,
ylabel={Running time (seconds)},
axis background/.style={fill=white},
legend pos=north west,
legend style={legend cell align=left,align=left,draw=black}
]
\addplot [color=black,dashed,line width=2.5pt]
  table[row sep=crcr]{%
20	0.032232064377\\
23	0.0322225516573333\\
26	0.032320196787\\
30	0.03253724499\\
35	0.0327168425404\\
40	0.03284850167\\
45	0.0329342725126\\
52	0.0328274117724\\
60	0.0325530805918\\
68	0.032260553497\\
78	0.0319638942718\\
90	0.0316604323194\\
103	0.0313996016812\\
118	0.0311839647636\\
135	0.0309938668008\\
155	0.0308485636622\\
177	0.0307455065534\\
203	0.0306136090404\\
233	0.0305288034864\\
267	0.0305096739652\\
306	0.0305485868684\\
351	0.0307236340786\\
402	0.0310413930128\\
461	0.0315462419028\\
528	0.0320597441924\\
606	0.0327264456868\\
694	0.033449956967\\
796	0.0343506600076\\
912	0.0352978171568\\
1045	0.0364344222752\\
1198	0.0377114253934\\
1373	0.0392445791364\\
1574	0.0411373117398\\
1804	0.0431856225624\\
2067	0.0456631620304\\
2370	0.0485998519992\\
2716	0.051868446325\\
3113	0.0556447749244\\
3568	0.0601561919062\\
4089	0.0653470329238\\
4687	0.0715249357112\\
5372	0.0788790499278\\
6157	0.087271745781\\
7057	0.096900168264\\
8089	0.1070327792892\\
9271	0.118684095759\\
10626	0.1316654867478\\
12179	0.146440707807\\
13960	0.161906671637667\\
16000	0.180591975895\\
};
\addlegendentry{FP2};

\addplot [color=black!50!green,loosely dashed,line width=2.5pt]
  table[row sep=crcr]{%
20	0.202278183875\\
23	0.192437046203667\\
26	0.1864855473484\\
30	0.1807608321088\\
35	0.1769778121018\\
40	0.1738953971764\\
45	0.1719674634826\\
52	0.1695667590738\\
60	0.1676692372698\\
68	0.1654322501982\\
78	0.1641538485868\\
90	0.1620850292872\\
103	0.1608074709882\\
118	0.1593667911764\\
135	0.158741615439\\
155	0.1570396107812\\
177	0.1559428548044\\
203	0.1551228816396\\
233	0.1543948671164\\
267	0.1533302022834\\
306	0.1536156482646\\
351	0.1537792535728\\
402	0.1538943149966\\
461	0.154817381937\\
528	0.1555330036816\\
606	0.1568042752896\\
694	0.1586185466286\\
796	0.1607811467934\\
912	0.1628367523974\\
1045	0.1652491302034\\
1198	0.1671403113984\\
1373	0.1701059108366\\
1574	0.1740880669666\\
1804	0.1781821132114\\
2067	0.1836075723838\\
2370	0.1898692268548\\
2716	0.1965266619402\\
3113	0.2037598522238\\
3568	0.212280745638\\
4089	0.2219128306482\\
4687	0.2333275679046\\
5372	0.2471153115178\\
6157	0.2618922586646\\
7057	0.2793437784928\\
8089	0.2996876713666\\
9271	0.3230670112676\\
10626	0.3483601840228\\
12179	0.3793069759214\\
13960	0.410925449619333\\
16000	0.450645738422001\\
};
\addlegendentry{LBFGS};

\addplot [color=red,solid,line width=2.5pt]
  table[row sep=crcr]{%
20	0.0875482299780001\\
23	0.0853458308573334\\
26	0.0847047133202\\
30	0.0840333643316\\
35	0.0836740572264\\
40	0.0836140227092\\
45	0.0837374756918\\
52	0.0830529415488001\\
60	0.0826947566544\\
68	0.0823738951854\\
78	0.0820443478108\\
90	0.0816899440934\\
103	0.081636987575\\
118	0.0815431985306\\
135	0.081670293786\\
155	0.0818981539358\\
177	0.0822281717034\\
203	0.0825269655378001\\
233	0.0828539913760001\\
267	0.083343667282\\
306	0.0840109680052\\
351	0.0849187941478\\
402	0.0860581478658\\
461	0.0878287688286\\
528	0.0895228637294\\
606	0.091555223786\\
694	0.0938473285172\\
796	0.0966881654354\\
912	0.0996429820542\\
1045	0.1030420575042\\
1198	0.1068928312132\\
1373	0.1115248762918\\
1574	0.1168976631898\\
1804	0.1226508055384\\
2067	0.1295438168472\\
2370	0.1374739218908\\
2716	0.1462419974878\\
3113	0.156532325653\\
3568	0.1681601660808\\
4089	0.1812984452666\\
4687	0.1964184082072\\
5372	0.2143576227442\\
6157	0.234135871043\\
7057	0.2574409195148\\
8089	0.2831951110914\\
9271	0.3124335235882\\
10626	0.3447221178032\\
12179	0.382357851075\\
13960	0.420695734962667\\
16000	0.468507117407001\\
};
\addlegendentry{CG};

\addplot [color=red!50!green,solid,line width=2.5pt]
  table[row sep=crcr]{%
20	0.156463534893\\
23	0.154368754946\\
26	0.1548065482206\\
30	0.1550110604516\\
35	0.1555971434594\\
40	0.156127756835\\
45	0.156598928396\\
52	0.1549723216006\\
60	0.1534446804418\\
68	0.1518518667912\\
78	0.1503674615088\\
90	0.148982032964\\
103	0.1481303939592\\
118	0.147723350384\\
135	0.1477257218178\\
155	0.1479610803254\\
177	0.148709860465\\
203	0.14951754073\\
233	0.1505413428156\\
267	0.151872885335\\
306	0.15359394795\\
351	0.1555802536366\\
402	0.1583393244666\\
461	0.16228705005\\
528	0.1660955337478\\
606	0.1709030698574\\
694	0.1762192968216\\
796	0.1825757369572\\
912	0.1891319450594\\
1045	0.19709107555\\
1198	0.2061057848916\\
1373	0.2169398580564\\
1574	0.2294488578032\\
1804	0.2429623438326\\
2067	0.2593935127022\\
2370	0.2782060308982\\
2716	0.2992631728748\\
3113	0.3238078044374\\
3568	0.3522468648338\\
4089	0.3842855837252\\
4687	0.4209908953918\\
5372	0.4634182498944\\
6157	0.5116683004486\\
7057	0.566853679427\\
8089	0.6307107590298\\
9271	0.702655239406601\\
10626	0.784557765022801\\
12179	0.879489056790401\\
13960	0.976803869897334\\
16000	1.098090835062\\
};
\addlegendentry{TR};

\end{axis}
\end{tikzpicture}%
}
\caption{\small Effect of number of data points on the convergence speed.}
\label{fig.fnSam}
\end{figure}
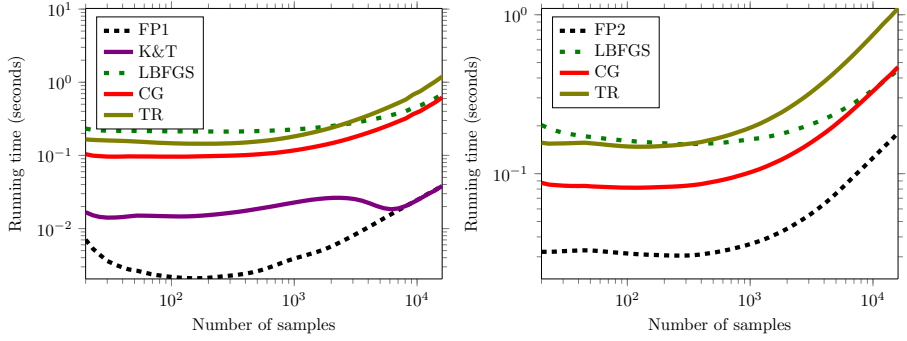

In Figures~\ref{fig.faRnd}-\ref{fig.fnSam}, we investigate the effect of different parameters on the convergence behavior. In all these figures, the optimization algorithm stops when the difference of average log-likelihood in two consecutive steps falls below $10^{-6}$. The plots are averaged over 1,000 different runs of algorithms. For each run, we sample data points from an EG distribution with a random scatter matrix.

To investigate the effect of shape parameter on the convergence, we report the convergence speed as a function of the shape parameter in Fig.~\ref{fig.faRnd}. The X-axis for the nonconcave case (left plot) is chosen to be $2a/q$, where the dimensionality is $q=16$. The X-axis for the concave case is $a-q/2$. In both plots in this figure, the sample size is $n=1000$.

It can be seen in Fig.~\ref{fig.faRnd} that for values of $a$ closed to $q/2$, i.e. when the distribution is closer to the Gaussian distribution, the Kent-Tyler method outperforms our fixed-point iteration, while for smaller values of $a$, our proposed method works better than Kent-Tyler. This shows that the method of choosing scale parameter mentioned in Remark~\ref{rmk.1} works well for smaller values of $a$.

A careful implementation of an optimization algorithm involves finding a good initial point. For computing the scatter matrix of an EGD, a possible good candidate is the sample covariance matrix. The results of optimization when we use the sample covariance matrix as initialization is shown in Fig.~\ref{fig.faSam}. The only difference between this figure and Fig.~\ref{fig.faRnd} is in the initialization. Apparently, our proposed fixed-point algorithm benefits the most. With this initialization, our proposed fixed-point algorithm always performs equally well or outperforms the Kent-Tyler method for all values of $a$.

In another experiment, we investigate the effect of dimensionality on the performance of different methods. The result is shown in Fig.~\ref{fig.fdSam}. For the nonconcave case, the shape parameter is $a=q/2/20$ and for the concave case, the shape parameter is $a=q/2+20$. The number of data points for different dimensions is $n=100q$. 
In general, we observed that when the shape parameter is chosen to be a fraction of $q/2$ for the nonconcave case and a constant addition to $q/2$ for the concave case and when the number of data points increase linearly by increasing dimensionality, the relative performance of different 
optimization methods stays almost equal for different dimensionality.

Since increasing the number of data points improves the accuracy of the initial estimate, we investigate the effect of number of data points in Fig.~\ref{fig.fnSam}. The dimensionality is equal to $q=16$ and the shape parameter for the nonconcave case is $a=q/2/20$ and for the concave case is $a=q/2+20$. For the concave case, relative performance of different optimization methods are almost equal for different number of data points. Except LBFGS, whose performance improves mainly due to amortization of  the overhead needed in addition to the computation of the function and gradient. For the nonconcave case, performance of our fixed-point method is better than the other methods. The performance of our method degrades for very small number of data points (on order of $n=10q$), because the sample covariance is not a very accurate initial estimate for smaller number of data points. For very large number of data points, the performance of Kent-Tyler method reaches the performance of our fixed-point method, because the initial estimate is very close to the optimum.

\section{Mixture modeling with EGDs}
\label{sec:mixture}
After presenting our results on ML estimation, we are now ready to discuss mixture modeling using EGDs. 
A $K$-component mixture of EGDs (MEG) has the density
\begin{equation}
  \label{eq:4}
  p(\vx)=\nlsum_{k=1}^K p_k p_{eg}(\vx;\msigma_k,a_k,b_k),\qquad\text{where}\ \sum\nolimits_kp_k=1,\  p_k \ge 0.
\end{equation}

We use block-coordinate ascent for maximizing the mixture log-likelihood. Specifically, we fix $a_k$ and $b_k$ and apply one step of EM to obtain $\msigma_k$ ($1\le k \le K$) using our fixed-point ML algorithms. Then, we fix $\msigma_k$, and estimate $a_k$, $b_k$. Here, the following variable change $\upsilon_k=\vx^T \msigma_k \vx$ proves helpful, because with it the density~\eqref{eq:4} turns into
\begin{equation*}
  p(\vx)=\nlsum_{k=1}^K p_k p_{ga}(\upsilon_k;a_{k},b_{k} ),
\end{equation*}
where $p_{ga}$ is the gamma density~\eqref{eqn_2a}.

The two main steps of an EM algorithm for the first stage are as follows:
\begin{list}{$\bullet$}{\leftmargin=1em}
\item
\emph{E-step}: For each data-point $i$ and component $k$, compute the following weights:
\begin{eqnarray*}
t_{ki} = \frac{p_k p_{eg}(\vx_i;\msigma_k,a_k,b_k)}{\sum_{l=1}^K p_l p_{eg}(\vx_i;\msigma_l,a_l,b_l)}= \frac{p_kp_{ga}(\upsilon_{ki};a_{k},b_{k} )}{\sum_{l=1}^K p_l p_{ga}(\upsilon_{ki};a_{l},b_{l} )}.
\end{eqnarray*}
\item
\emph{M-step}: Update the scatter matrices by maximizing the weighted log-likelihoods:
\begin{eqnarray*}
\ell_k(\msigma_k,a_k,b_k;\{\vx_i\}_{i=1}^n) = \sum_{i=1}^n t_{ki} \log p_{eg}(\vx_i;\msigma_k,a_k,b_k).
\end{eqnarray*}
\noindent The component probabilities  $p_k$ are updated as usual $p_k =n^{-1}\sum_{i=1}^n t_{ki}$.
\end{list}
The fixed-point methods of Section~\ref{sec:mle} can be easily modified to accommodate weighted log-likelihoods. 

Similar to the first stage, one step of EM for the second stage also consists of two steps that are applied sequentially until convergence. The E-step and updates to $p_k$ are similar to the first stage. But for updating the $a_k$ and $b_k$ parameters in the M-step, we maximize the following objective function:
\begin{align*}
\ell_k(a_{k},b_{k};\{\upsilon_{ki}\}_{i=1}^n)=\nlsum_{i=1}^{n} t_{ki}  \log p_{ga}(\upsilon_{ki}|a_k,b_k).
\end{align*}

The maximum weighted log-likelihood estimates of these parameters can be calculated efficiently using a generalized Newton method \citep{minka02}. Modifying the method explained in \citep{minka02} to account for weights, we obtain the following fixed-point iteration:
\begin{equation*}
\frac{1}{a_{k_{new}}}=\frac{1}{a_k}+\frac{\overline{\log \upsilon_k}- \log \bar{\upsilon}_k + \log a_k - \Psi (a_k)}{a_k^2 (\frac{1}{a_k}- \Psi' (a_k))},
\end{equation*}
where $\bar{z}$ is the weighted mean over $z$ ($\sum_{i}{t_{ki} z_{ki}}/\sum_{i}t_{ki}$)
 and $\Psi$ is the digamma function. The other parameter is calculated simply using the equation
\begin{equation*}
b_k=\bar{\upsilon}_k / a_k.
\end{equation*}

\section{Application: statistics of natural images}
A natural image dataset contains out-door images in a rural environment taken mostly from nature and landscape. It is in contrast with artificial images, like paintings, in-door images, etc., where most objects in the images are man-made. By natural image statistics, we mean finding a probability density function for natural image data. Equivalently, it means modeling the regularities and redundancies in natural image data. A non-regular image would be a random image where all its pixel values are independent. Natural images are far from random and show a wide range of regularities like textures, objects, etc. An accurate model for the statistics of natural images would be valuable for computational neuroscience studies and modeling the visual pathway~\citep{simoncelli2001natural}.
Having an accurate estimate of image densities is also important in many computer vision applications like compression~\citep{bethge2014method}, denoising~\citep{zoran12} and many other applications.

\begin{figure}
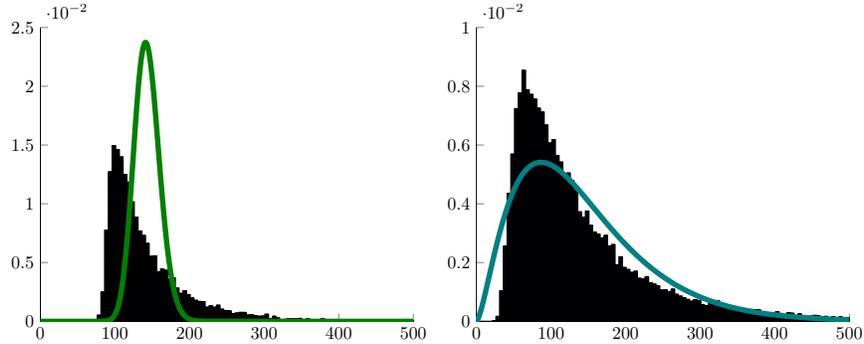

  \centering
  \resizebox{.43\textwidth}{!}{\input{histmvn.tikz}}
  \resizebox{.43\textwidth}{!}{\input{histeg.tikz}}
  \caption{Plots of the fitted radial density of one component randomly chosen from 8 components in a mixture of Gaussians (left) and in MEG (right). The overlap between fitted density and histogram is 86\% for the case of MEG and 46\% for the case of mixtures of Gaussians. }
  \label{fig:egvsgauss}
\end{figure}

We use MEG to model statistical distribution of natural image patches and compare its performance to some other models. 
The data used for fitting the models are patches sampled from random locations in a natural image dataset. Fig.~\ref{fig:egvsgauss} provides intuition as to why we model the statistics of image patches using MEGs rather than just a mixture of Gaussians.

\begin{figure}
 \includegraphics[width=.8\textwidth]{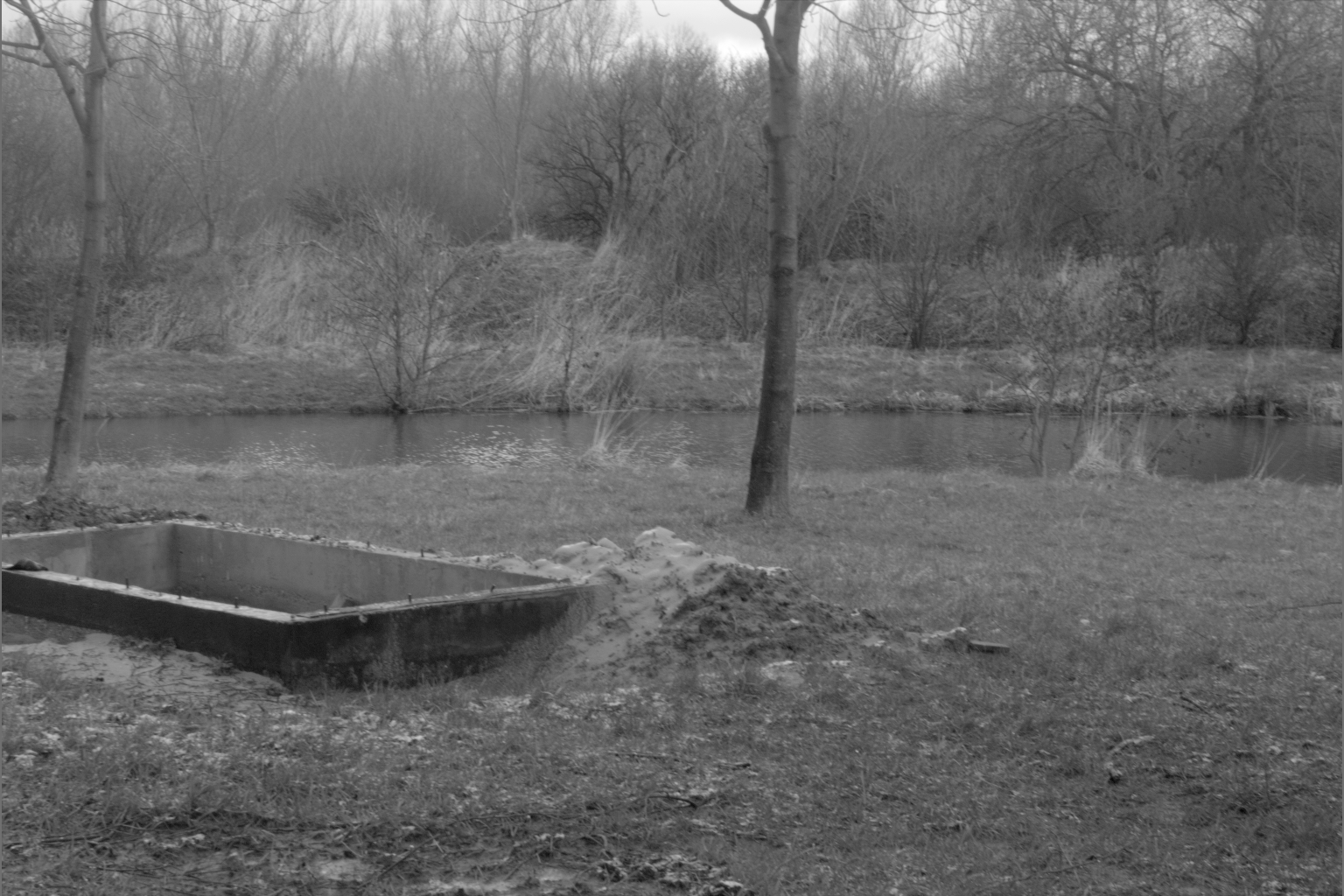}
\caption{\small A typical image chosen from van Hateren dataset.}
\label{fig.hateren}
\end{figure}

We extracted image patches of two different sizes $6\times6$ and $12\times12$ from random locations in the van Hateren dataset \citep{vanhateren98}. This dataset contains images from a forest-like environment in Netherlands. A typical randomly chosen image of this dataset is shown in Fig.~\ref{fig.hateren}.  This dataset has been used extensively in many computer vision and neuroscience studies especially those studies involving how human visual pathway is adapted to the statistics of natural images~\citep{clark2014flies, maboudi2015representation}. This dataset contains 4167 images; we excluded images that had problems, e.g., were noisy, blurred, etc. We extracted 200,000 training image patches, and 10 sets of 20,000 test image patches from the remaining 3632 images. 

We preprocessed image patches by log-transforming pixel intensities. Then, we added Gaussian white noise with variance equal to 0.2 percent of the variance of pixel intensities. It is important to add noise to the data, otherwise due to quantization the log-likelihood becomes infinite. The log-likelihood values are sensitive to the amount of noise added to the images, and adding even a small amount of extra noise can substantially reduce the obtained log-likelihood values. \citet{hosseini2012natural} investigated the effect of noise level on the log-likelihood and suggested the log-likelihood/noise-level curve as a measure for evaluating different models. 
\begin{table}\small
\begin{tabular}{c|cc|cc}
\hline
 & \multicolumn{2}{c|}{$6\times 6$} & \multicolumn{2}{c}{$12\times 12$}\\
 Model  & MI rate &  Parameters       & MI rate & Parameters \\
\hline
Gauss&2.41& 694 &2.53& 10468\\
GRBM&2.58&6696 &2.66&104544\\
DBN&2.60&39276 &2.69 & 623664\\
ICA&2.60& 1709&2.71&22337\\
EG&2.66& 660 &2.79 & 10326\\
HICA&2.71& 26909 &2.80&178493\\
MICA&2.77& 26924&2.85 & 178500\\
RG + ICA& 2.71&2340 &2.84 & 32634\\
MoG&2.77&10684 &2.84 & 83548\\
MEG&\textcolor{red}{\textbf{2.79}}&10140 &\textcolor{red}{\textbf{2.89}} & 81268\\
\hline
\end{tabular}
\vspace{0.45cm}
\caption{\small MI rate (bits/pixel; higher is better) and effective number of parameters for different models and two different patch sizes. The differences in MI rate attained are significant (please see text for discussion).}
\label{table_Comparison}
\end{table}

We evaluate the performance of different models using the \emph{multi-information rate} (MI rate) criterion. MI rate (in bits/pixel) measures the number of bits per pixel that one saves compressing the patch jointly compared to compressing all pixels independently. Formally, it is defined as
\begin{eqnarray*}
\text{MI rate} \approx \bigl ( H(X_0) + \tfrac{1}{n(q-1)}  \ell(\vtheta;\{\vx_i\}_{i=1}^n) \bigr )/ \log 2,
\end{eqnarray*}
where $H(X_0)$ is the entropy of one pixel and $q$ is the patch-size. The relation becomes exact if  $n \to \infty$ \citep{hosseini10}.

Table~\ref{table_Comparison} summarizes the performance of different procedures using MI rate.\footnote{Except DBN and GRBM, all other models were trained using our toolbox for mixture modeling available at: \href{http://visionlab.ut.ac.it/mixest}{http://visionlab.ut.ac.it/mixest} } The numerical values reported had very small error bars (variance) between 0.004--0.006, so we do not include these in the comparisons to avoid clutter.  For all models except the Gaussian restricted Boltzmann machine (GRBM) and the deep belief network (DBN), the DC component is modeled independently using a mixture of Gaussians with 10 components.  Two different patch sizes are included in order to observe how the MI rate estimates of different models change if the patch size increases. Among the different methods, MEG shows the best performance, \emph{yielding the highest MI rate per pixel.}

In the table, Gauss denotes the simple Gaussian model; the MI rate captured by this model is called the amount of second-order information present in the data. RG+ICA corresponds to radial Gaussianisation followed by one layer of independent component analysis (ICA)~\citep{hosseini09}. The number of layers in hierarchical ICA (HICA) \citep{hosseini09} and the number of components in MoG (mixtures of Gaussians) \citep{zoran12}, MEG and MICA (mixtures of ICAs)~\citep{mehrjou15} is 16 for $6\times6$ patches and 8 for $12\times12$ patches. Note that models like MoG, HICA and MICA are universal approximators, therefore theoretically they may reach the performance of MEG \emph{but with more parameters}. In practice, however,  parsimonious models are usually preferred. The MI rate of DBN and GRBM were evaluated by the method explained in \citep{theis11}. Similar to~\citep{theis11}, we also observed that increasing the number of layers beyond two layers only worsens the results for DBN. The number of hidden variables for GRBM and for both layers in DBN are 144 for $6\times6$ and 720 for $12\times12$ patches.

We emphasize that \emph{the differences in MI rate shown in Table~\ref{table_Comparison} are significant}, because closer to the upper limit of the MI rate any improvement means capturing a lot of perceptually relevant regularities of the underlying distribution, a claim grounded in the recent  psychophysical results in \citep{gerhard13}. 

\begin{figure}
 \includegraphics[width=0.32\textwidth]{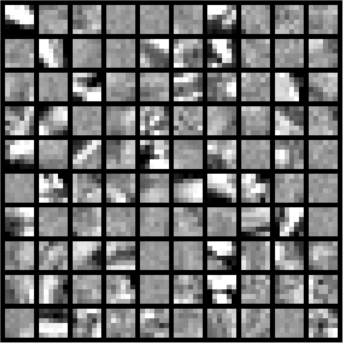}
 \includegraphics[width=0.32\textwidth]{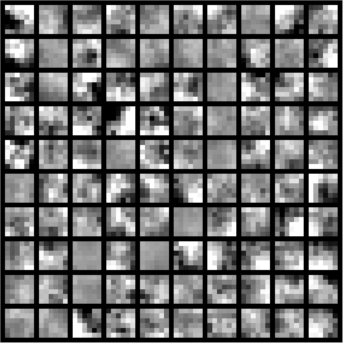}
  \includegraphics[width=0.32\textwidth]{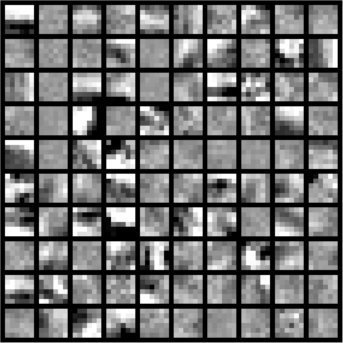}
\caption{\small From Left-to-right: natural image samples, samples from EG model, samples from MEG with 16 components. There are total number of 100 samples of size $6\times6$ that are organized in a 10 by 10 grid.}
\label{fig.samples}
\end{figure}

\begin{figure}
 \includegraphics[width=0.32\textwidth]{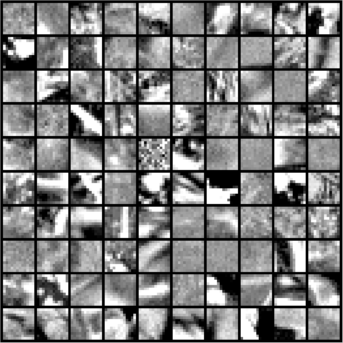}
 \includegraphics[width=0.32\textwidth]{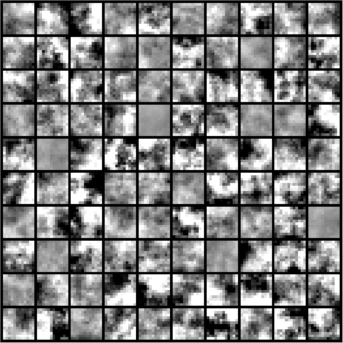}
  \includegraphics[width=0.32\textwidth]{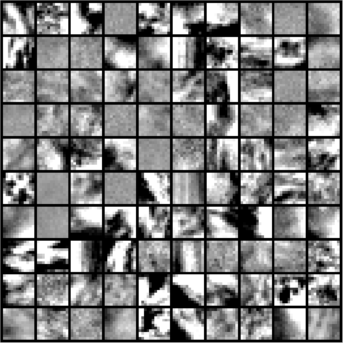}
\caption{\small The same results as Fig.~\ref{fig.samples} for $12\times12$ patches.}
\label{fig.samples2}
\end{figure}

To visualize how better MI rate corresponds to capturing more regularities. We sample image patches from two different models, the EG distribution and the MEG distribution with 16 components. The result is shown in Fig.~\ref{fig.samples}, where middle and right images correspond to sample patches from EG and MEG models, respectively. The left image consists of some random image patches taken from the van Hateren dataset. Fig.~\ref{fig.samples2} is the same result as Fig.~\ref{fig.samples} but for patch sizes $12 \times 12$. In Fig.~\ref{fig.samples}, image patches sampled from MEG is almost indistinguishable from natural image patches. For $12 \times 12$ patch sizes, although MEG captured a lot of redundancy but it has not captured all regularities and samples are distinguishable from natural images.

\begin{figure}
\centering
 \resizebox{.75\textwidth}{!}{% This file was created by matlab2tikz v0.4.7 (commit 3442858e5a642c135c5e9dab6a960bee5b9c6f8d) running on MATLAB 7.8.
% Copyright (c) 2008--2014, Nico Schlömer <nico.schloemer@gmail.com>
% All rights reserved.
% Minimal pgfplots version: 1.3
% 
% The latest updates can be retrieved from
%   http://www.mathworks.com/matlabcentral/fileexchange/22022-matlab2tikz
% where you can also make suggestions and rate matlab2tikz.
% 
%
% defining custom colors
\definecolor{mycolor1}{rgb}{1.00000,1.00000,0.90000}%
\begin{tikzpicture}

\begin{axis}[%
width=4.57555555555555in,
height=3.40555555555556in,
scale only axis,
xmin=0,
xmax=17,
xlabel={Number of Mixtures/Layers},
ymin=2.4,
ymax=2.8000001,
ylabel={MI rate},
legend style={at={(0.660565476190476,0.256184523809524)},anchor=south west,draw=black,fill=white,legend cell align=left}
]
\addplot [color=black!50!blue,solid,line width=3.0pt,mark size=3.3pt,mark=*,mark options={solid,fill=mycolor1}]
  table[row sep=crcr]{1	2.68496204439163\\
2	2.72591277069144\\
3	2.74641265734907\\
4	2.75333990252954\\
5	2.75991233379821\\
6	2.7662939256738\\
7	2.76951722128393\\
8	2.77213159406956\\
9	2.77522595000892\\
10	2.7770016513178\\
11	2.77888547995692\\
12	2.78133703277951\\
13	2.78280392527517\\
14	2.78422556352351\\
15	2.78577335384482\\
16	2.78621125143869\\
};
\addlegendentry{MEG};

\addplot [color=black!50!red,solid,line width=3.0pt,mark size=3.3pt,mark=*,mark options={solid,fill=mycolor1}]
  table[row sep=crcr]{1	2.40780500951037\\
2	2.63063580772686\\
3	2.68781403551295\\
4	2.71011532259705\\
5	2.72498971953096\\
6	2.73332702172871\\
7	2.74256443418034\\
8	2.74844503113872\\
9	2.75244339404264\\
10	2.75619339852785\\
11	2.75832851911806\\
12	2.76185380273416\\
13	2.76535255508431\\
14	2.76698300058356\\
15	2.76881414070956\\
16	2.77074154922038\\
};
\addlegendentry{MoG};

\addplot [color=black!50!green,dashed,line width=3.0pt,mark size=3.3pt,mark=*,mark options={solid,fill=mycolor1}]
  table[row sep=crcr]{0	2.70820828051191\\
17	2.70820828051191\\
};
\addlegendentry{RG+ICA};

\addplot [color=red!50!green,solid,line width=3.0pt,mark size=3.3pt,mark=*,mark options={solid,fill=mycolor1}]
  table[row sep=crcr]{1	2.6011172937737\\
2	2.70250903803481\\
3	2.72434429819983\\
4	2.73888771340132\\
5	2.74701113174669\\
6	2.7548249400017\\
7	2.76028474188616\\
8	2.76360214940882\\
9	2.76613812910464\\
10	2.76723679000961\\
11	2.76824478823675\\
12	2.76986440069444\\
13	2.77189993653157\\
14	2.77214911390051\\
15	2.7733291546181\\
16	2.77323434921832\\
};
\addlegendentry{MICA};

\addplot [color=teal,solid,line width=3.0pt,mark size=3.3pt,mark=*,mark options={solid,fill=mycolor1}]
  table[row sep=crcr]{1	2.6011172937737\\
2	2.65713698836075\\
3	2.67840367227761\\
4	2.68997829225181\\
5	2.69678886230593\\
6	2.70072651375408\\
7	2.70314092222048\\
8	2.70484540245966\\
9	2.70598306708727\\
10	2.70643208264278\\
11	2.70696986199532\\
12	2.70753730980613\\
13	2.70754108719085\\
14	2.70754076242206\\
15	2.70778472665717\\
16	2.7079204408622\\
};
\addlegendentry{HICA};

\addplot [color=violet,solid,line width=3.0pt,mark size=3.3pt,mark=*,mark options={solid,fill=mycolor1}]
  table[row sep=crcr]{1	2.57782284509471\\
2	2.60081940404648\\
};
\addlegendentry{DBN};

\addplot [color=black!50!blue,dashed,line width=3.0pt,mark size=3.3pt,mark=*,mark options={solid,fill=mycolor1}]
  table[row sep=crcr]{0	2.40780500951037\\
17	2.40780500951037\\
};
\addlegendentry{Gauss};

\end{axis}
\end{tikzpicture}%
}
\caption{\small MI rate for MEG and other methods by increasing number of parameters. Unsurprisingly, with large enough number of parameters (number of mixture components / layers) the differences between the models become less severe, but MEG still retains an edge.}
\label{fig.numMix}
\end{figure}
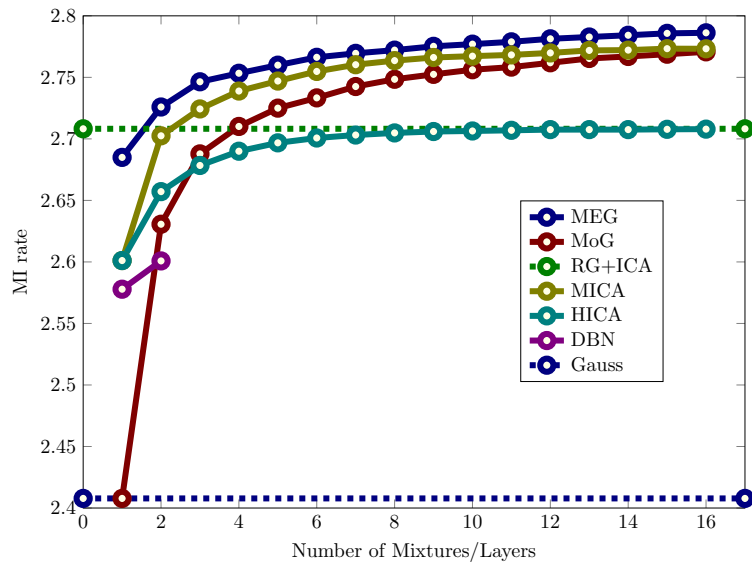

Finally, Fig.~\ref{fig.numMix} visualizes the effect of number of mixture components and number of layers on the performance of different models for $6 \times 6$ image patches. The baseline Gaussian MI rate is plotted as a dotted line.

\section{Discussion and future work}

We studied a powerful class of symmetric distributions, namely, elliptical gamma distributions. We presented theory outlining existence and uniqueness of maximum likelihood estimators for EGDs and developed simple and computationally effective algorithms computing these.

Several avenues of further research remain open. The most important direction is to study robust subspace recovery and its applications~\citep{soltanolkotabi12}. Other potential directions involve developing mathematical tools to study stochastic processes based on EGDs, as well as to investigate other applications where non-Gaussian data can benefit from EGDs or their mixture models.
We hope that the theory and its practical application outlined in this paper encourage a wider study of non-Gaussian modeling with EGDs or more general ECDs.

\bibliographystyle{abbrvnat}

\begin{appendices}

\section{Showing EG can be expressed as a GSM}  
\label{app:gsm}
From properties of Laplace transform, we know that the inverse Laplace transform of the following function
\begin{equation*}
p^{-v}e^{-cp},\quad v>0
\end{equation*}
is equal to 
\begin{equation*}
w(t) =\left \{
\begin{matrix}
0,&0<t<c,\\
\frac{(t-c)^{v-1}}{\Gamma(v)},&t>c.
\end{matrix}\right.
\end{equation*}
Using the definition of Laplace transform, we obtain:
\begin{equation*}
p^{-v}e^{-ap}=\int_{0}^{\infty}{w(t)e^{-tp}dt}.
\end{equation*}
Now assume $v=q/2-a$, $c=2b^{-1}$ and $p=\tfrac12 \vx^\top \msigma^{-1} \vx$, then the left term in the equation above is the EG density given in~\eqref{eq:1}. After straightforward computations, we can write the EG density as a scale mixture of Gaussian densities:
\begin{align*}
&p_{eg}(\vx;\msigma,a,b) = C |\msigma|^{-\tfrac 12} \int_{2b^{-1}}^{\infty}{w(t)e^{-t\vx^\top \msigma^{-1} \vx}dt} \nonumber\\
&= \int_{2b^{-1}}^{\infty}{ \frac{2^{a}\Gamma(\tfrac q2)}{\Gamma(a)b^a} (t)^{-\tfrac q2}\frac{(t-2b^{-1})^{\tfrac q2-a-1}}{\Gamma(\tfrac q2-a)} (2\pi)^{-\tfrac q2}|t^{-1}\msigma|^{-\tfrac12} \exp \biggl (-\tfrac12 t\vx^\top \msigma^{-1} \vx \biggr ) dt} \nonumber\\
& = \int_{2b^{-1}}^{\infty} \frac{\Gamma(q/2)}{\Gamma(\tfrac q2-a)\Gamma(a)}
\frac{(t-2b^{-1})^{\tfrac q2-a-1} 2^{a}}{(t)^{\tfrac q2} b^a} p_{n}(\vx;o,t^{-1}\msigma)dt.
\end{align*}
Without loss of generality assume $b=2$ and use the change of variable $u=1/t$, we obtain:
\begin{align*}
p_{eg}(\vx;\msigma,a,2) = \int_{0}^{1} \frac{\Gamma(\tfrac q2)}{\Gamma(\tfrac q2-a)\Gamma(a)} (1-u)^{\tfrac q2-a-1} u^{a-1}p_{n}(\vx;o,u\msigma) du.
\end{align*}
Interestingly, the first term is beta density with parameters $(q/2-a,a)$:
\begin{equation*}
p_{eg}(\vx;\msigma,a,2) = \int_{0}^{1} p_{\beta}(u;\tfrac q2-a,a)p_{n}(\vx;0,u\msigma) du.
\end{equation*} 

\section{Uniqueness of the log-likelihood critical point}
\label{app:uniqueML}
\begin{proof} By the assumption $\mm(\msigma_1,c,d)=\mm(\msigma_2,c,d)$, we have:
\begin{align*}
\begin{split}
c&\sum_{i=1}^{n}{\frac{\msigma_1^{-1/2}\vx_i\vx_ i^\top \msigma_1^{-1/2}}{\vx _i^\top \msigma_1^{-1}\vx _i}}+d\sum_{i=1}^{n}{\msigma_1^{-1/2}\vx _i\vx_ i^\top \msigma_1^{-1/2}} \\
&= c\sum_{i=1}^{n}{\frac{\msigma_2^{-1/2}\vx _i\vx_ i^\top \msigma_2^{-1/2}}{\vx _i^\top \msigma_2^{-1}\vx _i}}+d\sum_{i=1}^{n}{\msigma_2^{-1/2}\vx _i\vx_ i^\top \msigma_2^{-1/2}}.
\end{split}
\end{align*}
Substituting $\vz_ i=\msigma_2^{-1/2}\vx _i$ and $\bm S=\msigma_1^{-1/2} \msigma_2^{1/2}$ in the previous equation, we obtain:

\begin{eqnarray}
c\sum_{i=1}^{n}{\frac{ \bm S\vz_ i \vz_ i^\top\bm S^\top }{\vz_ i^\top  \ \bm S^\top \bm S \ 
 \vz_ i} } + d \sum_{i=1}^{n}{ \msigma\vz_ i \vz_ i^\top\msigma^\top }= 
c\sum_{i=1}^{n}{\frac{ \vz_ i \vz_ i^\top }{\vz_ i^\top  \  \vz_ i} } + d \sum_{i=1}^{n}{ \vz_ i \vz_ i^\top },
\label{app.eqn_2f}
\end{eqnarray}

Let $\vu$ be a right eigenvector for $\bm S$ corresponding to the eigenvalue $\lambda$, then multiplying~\eqref{app.eqn_2f} from left by $\vu^\top$ and from right by $\vu$ and using the fact that the following equality holds for the eigenprojection:
\begin{eqnarray*}
\vu^\top\bm S\vz&=&\lambda \vu^\top\vz,
\end{eqnarray*}
we obtain:
\begin{eqnarray}
c \lambda^2\sum_{i=1}^{n}{\frac{\left ( \vu^\top \vz_ i\right )^2}{\vz_ i^\top  \bm S \bm S  \vz_ i} } + d\lambda^2 \sum_{i=1}^{n}{\left ( \vu^\top \vz_ i\right )^2}= 
c\sum_{i=1}^{n}{\frac{\left ( \vu^\top \vz_ i\right )^2}{\vz_ i^\top  \  \vz_ i} } + d \sum_{i=1}^{n}{\left ( \vu^\top \vz_ i\right )^2}.
\label{app.eqn_2i}
\end{eqnarray}

Using the fact that the product of two positive definite matrices has positive eigenvalues \cite[p.465]{horn85}, following two inequalities can be derived by straightforward computations:
\begin{eqnarray}
\lambda_q^2 \vz_ i^\top \vz_ i \leq \vz_ i^\top   \ms^\top \ms \vz_ i \leq \lambda_1^2 \vz_ i^\top \vz_ i,
\label{app.eqn_ine}
\end{eqnarray}
where $\lambda_1$ and $\lambda_q$ are the largest and the smallest eigenvalues respectively. It is clear that if $\lambda_1>1$ or $\lambda_q<1$ then inequalities in~\eqref{app.eqn_ine} contradicts the equality in~\eqref{app.eqn_2i}. Therefore, all eigenvalues of $\ms$ need to be equal to one which implies $\ms=\mi$ or $\msigma_1=\msigma_2$ and the proof is complete.
\color{black}
\end{proof}

\section{Proof of Lemma~\ref{lem:one}}
\label{sec:lemone}
\begin{proof}
By definition,
\begin{equation*}
\begin{split}
\mi&=\mn_p^{-1/2} \mn_p \mn_p^{-1/2}\\&= c \sum_{i=1}^{n}{\frac{\mn_p^{-1/2} \mgamma_p^{-1/2}\ \vy_i\vy_i^\top  \ \mgamma_p^{-1/2} \mn_p^{-1/2}}{\vy_i^\top  \ \mgamma_p^{-1} \  \vy_i} }  
+\mn_p^{-1/2}\mgamma_p^{-1/2}\mgamma_p^{-1/2}\mn_p^{-1/2}.
\end{split}
\end{equation*}
We multiply both the numerator and denominator of the first term by $\vy_i^\top   \mgamma_{p+1}^{-1} \vy_i$, and in the numerator we replace $\vy_i^\top   \mgamma_{p+1}^{-1} \vy_i$ by $\frac 1 \alpha_p \vy_i^\top   \mgamma_p^{-1/2} \mn_p^{-1} \mgamma_p^{-1/2} \vy_i$. In addition, we multiply on both sides by an orthogonal matrix $\mq_p$. This yields:
\begin{equation}
\begin{split}
\label{eqn_2j}
\mi =& \frac{c}{\alpha_p} \sum_{i=1}^{n}{\frac{\mq_p \mn_p^{-\frac 1 2} \mgamma_p^{-\frac 1 2}\ \vy_i\vy_i^\top  \ \mgamma_p^{-\frac{1}{ 2}} \mn_p^{-\frac 1 2} \mq_p^\top  }{\vy_i^\top  \ \mgamma_{p+1}^{-1} \  \vy_i} \frac{\vy_i^\top   \mgamma_p^{-\frac{1}{2}} \mn_p^{-1} \mgamma_p^{-\frac{1}{2}} \vy_i}{\vy_i^\top   \mgamma_{p}^{-1} \vy_i}}\\
&+\mq_p\mn_p^{-\frac 1 2}\mgamma_p^{-\frac 1 2}\mgamma_p^{-\frac 1 2}\mn_p^{-\frac 1 2} \mq_p. 
\end{split}
\end{equation} 
Since the square root of the matrix $\mgamma_p^{\frac{1}{2}} \mn_p \mgamma_p^{\frac{1}{2}}$ can be written as $\mgamma_p^{\frac{1}{2}} \mn_p^{\frac{1}{2}} \mq_p^\top$, using~\eqref{eqn_iter}, we obtain the identity
\begin{eqnarray}
\mq_p \mn_p^{-\frac 1 2} \mgamma_p^{-\frac 1 2}=\sqrt{\alpha_p}\mgamma_{p+1}^{-\frac 1 2}.
\label{eqn_2k}
\end{eqnarray}
Now substitute \eqref{eqn_2k} into~\eqref{eqn_2j} to obtain the equation
\begin{eqnarray}
\mi =c \sum_{i=1}^{n}{\frac{\mgamma_{p+1}^{-\frac{1}{2}}\ \vy_i^\top   \vy_i\ \mgamma_{p+1}^{-\frac{1}{2}}}{\vy_i^\top  \ \mgamma_{p+1}^{-1} \  \vy_i} \frac{\vy_i^\top   \mgamma_p^{-\frac{1}{2}} \mn_p^{-1} \mgamma_p^{-\frac{1}{2}} \vy_i}{\vy_i^\top   \mgamma_{p}^{-1} \vy_i}}+\alpha_p\mgamma_{p+1}^{-1}.
\label{eqn_2l} 
\end{eqnarray} 
By the extremal properties of the largest and smallest eigenvalues, we know that
\begin{eqnarray}
\lambda_{1,p}^{-1} \leq \frac{\vy_i^\top   \mgamma_p^{-\frac{1}{2}} \mn_p^{-1} \mgamma_p^{-\frac{1}{2}} \vy_i}{\vy_i^\top   \mgamma_{p}^{-1} \vy_i} \leq \lambda_{q,p}^{-1}.
\label{eqn_2m}   
\end{eqnarray} 
Therefore, on applying the inequalities~\eqref{eqn_2m} to~\eqref{eqn_2l}, we obtain following two inequalities:
\begin{align}
\label{eqn_2m1}
&\lambda_{q,p}^{-1}\left [ c \sum_{i=1}^{n}{\frac{\mgamma_{p+1}^{-\frac{1}{2}}\ \vy_i^\top   \vy_i\ \mgamma_{p+1}^{-\frac{1}{2}}}{\vy_i^\top  \ \mgamma_{p+1}^{-1} \  \vy_i}} \right ]+\alpha_p\mgamma_{p+1}^{-1} \geq \mi,\\
\label{eqn_2m2}   
&\mi \geq \lambda_{1,p}^{-1}\left [c \sum_{i=1}^{n}{\frac{\mgamma_{p+1}^{-\frac{1}{2}}\ \vy_i^\top   \vy_i\ \mgamma_{p+1}^{-\frac{1}{2}}}{\vy_i^\top  \ \mgamma_{p+1}^{-1} \  \vy_i}} \right ]+\alpha_p\mgamma_{p+1}^{-1}.
\end{align}
Rearranging the equality in~\eqref{eqn_2i2}, we have the equality
\begin{eqnarray}
c\sum_{i=1}^{n}{\frac{\mgamma_{p+1}^{-1/2}\vy_i\vy_i^\top  \mgamma_{p+1}^{-1/2}}{\vy_i^\top  \mgamma_{p+1}^{-1}\vy_i}}=\mn_{p+1}-\mgamma_{p+1}^{-1},
\label{eqn_2n}
\end{eqnarray} 
which can be applied to~\eqref{eqn_2m1} to obtain the following inequality:
\begin{eqnarray}
\lambda_{q,p}^{-1}\bigl[ \mn_{p+1}-\mgamma_{p+1}^{-1}\bigr ]+\alpha_p\mgamma_{p+1}^{-1} \geq \mi,
\label{eqn_2fpow}
\end{eqnarray}
which in turn can be rearranged to
\begin{eqnarray}
\mn_{p+1} \geq \lambda_{q,p}\mi + \mgamma_{p+1}^{-1}\left (1-\alpha_p\lambda_{q,p}\right ).
\label{eqn_2p}
\end{eqnarray}
Writing the singular value decomposition of $\mn_{p+1}$ as $\mU_{p+1}\mlambda_{p+1}\mU_{p+1}^\top  $, and  multiplying~\eqref{eqn_2p} from the left by $\mU_{p+1}^\top  $ and from the right by $\mU_{p+1}$, we obtain
\begin{eqnarray}
\mlambda_{p+1} \geq \lambda_{q,p}\mi+ \underbrace{\mU_{p+1}^\top  \mgamma_{p+1}^{-1}\mU_{p+1}\left (1-\alpha_p\lambda_{q,p}\right )}_{\Xi}.
\end{eqnarray} 
Let $\lambda_{q,p}\leq\alpha_p^{-1}$, then if the data points span $\reals^{q}$, the matrix
$\Xi$
is positive semidefinite, and in particular its diagonal elements are nonnegative. Consequently, all diagonal elements of $\mlambda_{p+1}$ are larger than or equal to $\lambda_{q,p}$. Therefore, if $\lambda_{q,p}\leq \alpha_p^{-1}$, then $\lambda_{q,p+1} \geq \lambda_{q,p}$ holds true.\\[5pt]
Applying the same procedure to the other inequality~\eqref{eqn_2m2}, we obtain
\begin{eqnarray}
\mlambda_{p+1} \leq \lambda_{1,p}\mi- \underbrace{\mU_{p+1}^\top  \mgamma_{p+1}^{-1}\mU_{p+1}\left (\alpha_p\lambda_{1,p}-1\right)}_{\Xi'}.
\end{eqnarray} 
Let $\lambda_{1,p}\geq\alpha_p^{-1}$, then if the data points span $\reals^{q}$, the matrix 
$\Xi'$
is positive semidefinite. Therefore, all its diagonal are nonnegative, whereby all diagonal elements of $\mlambda_{p+1}$ are smaller or equal to $\lambda_{1,p}\mi$. Therefore, if $\lambda_{1,p}\geq\alpha_p^{-1}$ then $\lambda_{1,p+1} \leq \lambda_{1,p}$ holds true.
\end{proof}

\end{appendices}

\end{document}